\documentclass[twoside,twocolumn,9pt]{article}
\usepackage{extsizes}
\usepackage[super,sort&compress,comma]{natbib} 
\usepackage[version=3]{mhchem}
\usepackage[left=1.5cm, right=1.5cm, top=1.785cm, bottom=2.0cm]{geometry}
\usepackage{balance}
\usepackage{mathptmx}
\usepackage{sectsty}
\usepackage{graphicx} 
\usepackage{lastpage}
\usepackage[format=plain,justification=justified,singlelinecheck=false,font={stretch=1.125,small,sf},labelfont=bf,labelsep=space]{caption}
\usepackage{float}
\usepackage{fancyhdr}
\usepackage{fnpos}
\usepackage[english]{babel}
\addto{\captionsenglish}{%
  
}
\usepackage{array}
\usepackage{droidsans}
\usepackage{charter}
\usepackage[T1]{fontenc}
\usepackage[usenames,dvipsnames]{xcolor}
\usepackage{setspace}
\usepackage[compact]{titlesec}
\usepackage{hyperref}

\usepackage{epstopdf}

\definecolor{cream}{RGB}{222,217,201}

\begin{document}

\pagestyle{fancy}
\thispagestyle{plain}
\fancypagestyle{plain}{
\renewcommand{\headrulewidth}{0pt}
}

\makeFNbottom
\makeatletter
\renewcommand\LARGE{\@setfontsize\LARGE{15pt}{17}}
\renewcommand\Large{\@setfontsize\Large{12pt}{14}}
\renewcommand\large{\@setfontsize\large{10pt}{12}}
\renewcommand\footnotesize{\@setfontsize\footnotesize{7pt}{10}}
\makeatother

\renewcommand{\thefootnote}{\fnsymbol{footnote}}
\renewcommand\footnoterule{\vspace*{1pt}%
\color{cream}\hrule width 3.5in height 0.4pt \color{black}\vspace*{5pt}} 
\setcounter{secnumdepth}{5}

\makeatletter 
\renewcommand\@biblabel[1]{#1}            
\renewcommand\@makefntext[1]%
{\noindent\makebox[0pt][r]{\@thefnmark\,}#1}
\makeatother 
\renewcommand{\figurename}{\small{Fig.}~}
\sectionfont{\sffamily\Large}
\subsectionfont{\normalsize}
\subsubsectionfont{\bf}
\setstretch{1.125} 
\setlength{\skip\footins}{0.8cm}
\setlength{\footnotesep}{0.25cm}
\setlength{\jot}{10pt}
\titlespacing*{\section}{0pt}{4pt}{4pt}
\titlespacing*{\subsection}{0pt}{15pt}{1pt}

\fancyfoot{}
\fancyfoot[LO,RE]{\vspace{-7.1pt}\includegraphics[height=9pt]{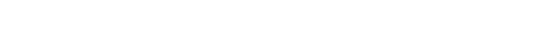}}
\fancyfoot[CO]{\vspace{-7.1pt}\hspace{13.2cm}\includegraphics{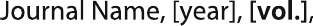}}
\fancyfoot[CE]{\vspace{-7.2pt}\hspace{-14.2cm}\includegraphics{head_foot/RF}}
\fancyfoot[RO]{\footnotesize{\sffamily{1--\pageref{LastPage} ~\textbar  \hspace{2pt}\thepage}}}
\fancyfoot[LE]{\footnotesize{\sffamily{\thepage~\textbar\hspace{3.45cm} 1--\pageref{LastPage}}}}
\fancyhead{}
\renewcommand{\headrulewidth}{0pt} 
\renewcommand{\footrulewidth}{0pt}
\setlength{\arrayrulewidth}{1pt}
\setlength{\columnsep}{6.5mm}
\setlength\bibsep{1pt}

\makeatletter 
\newlength{\figrulesep} 
\setlength{\figrulesep}{0.5\textfloatsep} 

\newcommand{\topfigrule}{\vspace*{-1pt}%
\noindent{\color{cream}\rule[-\figrulesep]{\columnwidth}{1.5pt}} }

\newcommand{\botfigrule}{\vspace*{-2pt}%
\noindent{\color{cream}\rule[\figrulesep]{\columnwidth}{1.5pt}} }

\newcommand{\dblfigrule}{\vspace*{-1pt}%
\noindent{\color{cream}\rule[-\figrulesep]{\textwidth}{1.5pt}} }

\makeatother

\twocolumn[
  \begin{@twocolumnfalse}
{\includegraphics[height=30pt]{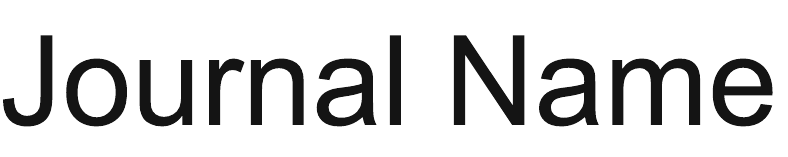}\hfill\raisebox{0pt}[0pt][0pt]{\includegraphics[height=55pt]{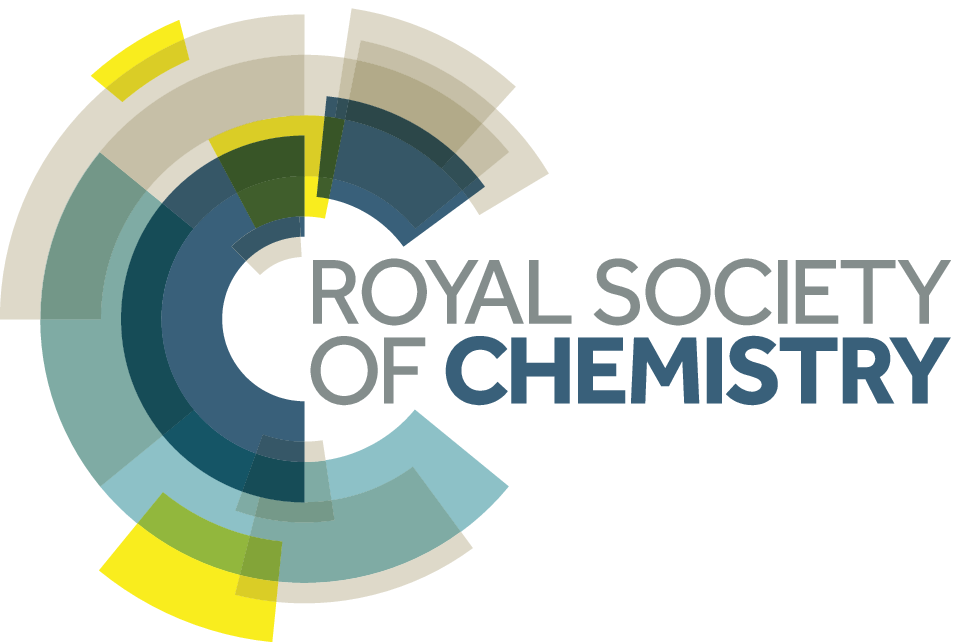}}\\[1ex]
\includegraphics[width=18.5cm]{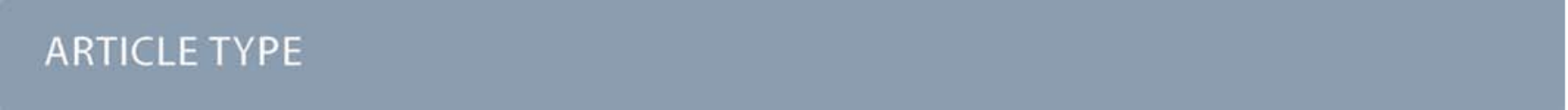}}\par
\vspace{1em}
\sffamily
\begin{tabular}{m{4.5cm} p{13.5cm} }

\includegraphics{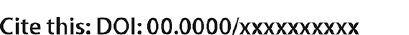} & \noindent\LARGE{\textbf{A theory of skyrmion crystal formation$^\dag$}} \\
\vspace{0.3cm} & \vspace{0.3cm} \\

 & \noindent\large{Xu-Chong Hu,\textit{$^{a,b}$} Hai-Tao Wu\textit{$^{a,b}$} and X. R. Wang\textit{$^{*a,b}$} } \\

\includegraphics{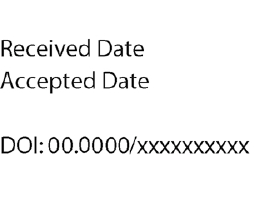} & \noindent{A generic theory about 
	skyrmion crystal (SkX) formation in chiral magnetic thin films and its 
	fascinating thermodynamic behaviours is presented. A chiral magnetic film 
	can have many metastable states with an arbitrary skyrmion density up 
	to a maximal value when a parameter $\kappa$, which measures the relative 
	Dzyaloshinskii-Moriya interaction (DMI) strength, is large enough. The 
	lowest energy state of an infinite film is a long zig-zagged ramified stripe 
	skyrmion occupied the whole film in the absence of a magnetic field. 
	Under an intermediate field perpendicular to the film, the lowest energy state 
	has a finite skyrmion density. This is why a chiral magnetic film is often in 
	a stripy state at a low field and a SkX only around an optimal field when $\kappa$ 
	is above a critical value. The lowest energy state is still a stripy helical state 
	no matter with or without a field when $\kappa$ is below the critical value. 
	The multi-metastable states explains the thermodynamic path dependences of 
	various metastable states of a film. Decrease of $\kappa$ value with the 
	temperature explains why SkXs become metastable at low temperature in many 
	skyrmion systems. These findings open a new avenue for SkX manipulation and 
	skyrmion-based applications.} \\
\end{tabular}

 \end{@twocolumnfalse} \vspace{0.6cm}

  ]

\renewcommand*\rmdefault{bch}\normalfont\upshape
\rmfamily
\section*{}
\vspace{-1cm}


\footnotetext{\textit{$^{a}$~Physics Department, The Hong Kong University of Science 
			and Technology (HKUST), Clear Water Bay, Kowloon, Hong Kong,}}
\footnotetext{\textit{$^{b}$~HKUST Shenzhen Research Institute, Shenzhen 518057, China}}
\footnotetext{\textit{$^{*}$~E-mail:phxwan@ust.hk }}
\footnotetext{\dag~Electronic Supplementary Information (ESI) available: Detail of the 
supporting videos. See DOI: 00.0000/00000000.}



\section{Introduction}

Skyrmion crystals (SkXs) have attracted enormous attention in the past 
decade because of their academic interest and potential applications 
\cite{roadmap,Klaui} since they were first unambiguously observed 
in chiral magnets \cite{Muhlbauer2009,Yu2010,Heinze,Romming}. 
Skyrmions and SkXs are good platform for studying various physics 
\cite{Pfleiderer1,the1,Pfleiderer3,ong,roadmap,Yuan2018,gong_prb_2020} such as 
the emergent electromagnetic fields and topological Hall effect \cite{the1,ong,
	Pfleiderer3,roadmap}. Three types of circular skyrmions in chiral magnets \cite
{roadmap,Bogdanov2001,Rossler2006,Yu2010,Kezsmarki2015,Nayak2017} have been 
identified, namely, Bloch skyrmions, hedgehog skyrmions, and anti-skyrmions. 
SkXs have been observed in many magnetic materials with Dzyaloshinskii-Moriya 
interaction (DMI) \cite{Yu2010,Muhlbauer2009,Heinze,Romming,Kezsmarki2015,
	Nayak2017} or with geometric frustration \cite{Okubo,Leonov,Kurumaji2019}.

Although many theories \cite{Leonov2016,Buhrandt2013,DOYI2009,han2010,Rybakov2016} 
predicted SkXs at low or even zero temperatures and in the absence of a magnetic field, 
it is an experimental fact \cite{skm-form-prm,Yu2011} that SkXs form thermodynamically 
under the assistance of an optimal magnetic field and near the Curie temperature to date.
Once formed, however, SkXs can be metastable in very large temperature-field regions. 
For example, a SkX can exist at zero magnetic field and a field much higher than 
an optimal value by cooling a SkX first in the optimal field to a low temperature 
followed by removal or rise of the magnetic field. A SkX disappears during the 
zero field warming and high field warming \cite{skm-form-prm,Tokura-nm16,roadmap}. 
A SkX is a thermal equilibrium state only in a narrow magnetic-field range and 
near the Curie temperature. The phase region for stable SkXs is normally larger 
in a thinner film \cite{Yu2011,Han2020,Yu2015} than that of a thicker film or a 
bulk material \cite{skm-form-prm,Muhlbauer2009}. Outside the range, SkXs can only 
be metastable states \cite{Leonov2016,Rybakov2016}. 
At zero field and a temperature much lower than the Curie temperature, the thermal 
equilibrium phase becomes a collection of stripe spin textures known as a helical 
state \cite{Thiaville2013,Tokura2006}. Existing theories \cite{Leonov2016,han2010,
	Buhrandt2013,DOYI2009,Rybakov2016} have not provided a satisfactory answer to the 
field effects and the thermodynamic path dependences of various experimentally 
observed phases. 

The thermal equilibrium state of a system results from the competition between 
the internal energy and entropy. Entropy dominates the higher temperature phases 
while the internal energy determines the lower-temperature ones. If one views a 
long flexible stripe as a long flexible polymer, then stripe entropy, according 
to de Gennes theory \cite{de-Gennes}, is proportional to the logarithm of stripe 
length: The end-end distribution function of a flexible polymer of $N$ monomers  
is $p(\vec x)\simeq [3/(2\pi N a^2)^{3/2}exp[-3\vec{x}^2/(2Na^2)]$, where $a$ is 
monomer size. In information theory, it is well known that the Gaussian 
probability distribution with a fixed mean and variance gives the maximal entropy. 
The polymer entropy is $S=-\int p\log(p) d^3\vec x\simeq \frac{3}{2}\log(N)$. 
Previous study \cite{Muhlbauer2009}, under the assumption of infinite long rigid 
stripes, argued that the translational entropy of helical states is much smaller 
than SkX entropy that is proportional to the logarithm of number of sites in one 
unit cell of the SkX. Obviously, real stripes are flexible as shown in 
both micromagnetic simulations and experiments and rigid assumption is incorrect. 
In reality, translational entropy cannot compete with the entropy of very long 
stripes in helical states mentioned. 
This view is consistent with experimental facts that numerous stripe structures, 
including ramified stripes and maze, were observed while few variations of SkX 
structures are possible \cite{Romming,Wang2020,legrand2017,Yu2011}. One interesting 
mystery about low entropy SkX is its metastability at lower temperatures even 
when a SkX is a thermal equilibrium state at higher temperatures \cite{roadmap}. 
This seemingly contradicts to the general principle that a higher temperature 
prefers a higher entropy state. Thus, a proper understanding is needed. 

In this paper, the roles of magnetic field in skyrmion crystal (SkX) formation 
is revealed. We show that a chiral magnetic thin film of a given size with 
an arbitrary number of skyrmions up to a critical value is metastable when 
skyrmion formation energy is negative and skyrmions are stripes that are usually 
called helical states. The energy and morphology of theses metastable states 
depend on the skyrmion density and the magnetic field perpendicular to the film. 
At zero field, the energy increases with skyrmion number or skyrmion density. 
Thus, the film at thermal equilibrium below the Curie temperature and at 
zero field should be helical states consisting of a few stripe skyrmions.
At non-zero field, the film with $Q_{\rm m}$ skyrmions has the lowest energy. 
$Q_{\rm m}$ first increases with the field up to an optimal value then decreases 
with the field. A parameter called $\kappa$ that measures the relative DMI 
strength to the exchange stiffness and magnetic anisotropy plays a crucial role. 
For $\kappa$ above a critical value, $Q_{\rm m}$ near the optimal field is large 
enough such that the average distance between two neighbouring skyrmions is 
comparable with skyrmion stripe width, and skyrmions form a SkX. 
For $\kappa$ below the critical value, the system prefers a stripy phase or a mixing 
phase consisting of stripes and circular skyrmions even in the optimal fields.\\

\section{Model and methodology} 

We consider a chiral magnetic thin film of thickness $d$ in $xy$-plane. 
The magnetic energy of a spin structure $\hat{m}$ with an interfacial (itf.) DMI is 
\begin{equation}
	\begin{aligned}
		E=&d\iint \lbrace A|\nabla \hat{m}|^2+D[(\hat{m}\cdot\nabla)m_z-m_z\nabla\cdot\hat{m}]
		\\&+K_{\rm u}(1-m_z^2)-\mu_0 M_{\rm s}\vec H_{\rm d}\cdot\hat{m}+\mu_0 H M_{\rm s}(1-m_z)\rbrace\,\mathrm{d}S,
	\end{aligned}
	\label{energy1}
\end{equation}
and energy with a bulk DMI is
\begin{equation}
	\begin{aligned}
		E=&d\iint \lbrace A|\nabla \hat{m}|^2+D\left[\hat{m}\cdot(\nabla\times\hat{m})\right]+
		K_{\rm u}(1-m_z^2)\\& -\mu_0 M_{\rm s}\vec H_{\rm d}\cdot\hat{m}+\mu_0 H M_{\rm s}(1-m_z)\rbrace\,\mathrm{d}S.
	\end{aligned}
	\label{energy2}
\end{equation}
$A$, $D$, $K_{\rm u}$, $H$, $M_{\rm s}$, $\vec{H}_{\rm d}$ and $\mu_0$ are exchange 
stiffness constant, DMI coefficient, the magneto-crystalline anisotropy, perpendicular 
magnetic field, the saturation magnetization, the demagnetizing field and the vacuum 
permeability, respectively. Ferromagnetic state of $m_z=1$ is set as zero energy $E=0$. 
For an ultra thin film, demagnetization effect can be included in the effective 
anisotropy $K=K_{\rm u}-\mu_0M_{\rm s}^2/2$. This is a good approximation when 
the film thickness $d$ is much smaller than the exchange length \cite{Xiansi}. 
It is known that isolated circular skyrmions are metastable state of energy 
$8\pi Ad\sqrt{1-\kappa}$ when $\kappa=\pi^2 D^2/(16AK) <1$ \cite{Xiansi}. 

Spin dynamics in a magnetic field is governed by the Landau-Lifshitz-Gilbert 
(LLG) equation,
\begin{equation}
	\frac{\partial \vec m}{\partial t} =-\gamma\vec m \times \vec H_{\rm eff} +
	\alpha \vec m \times \frac{\partial \vec m}{\partial t}, 
	\label{llg}
\end{equation}
where $\gamma$ and $\alpha$ are respectively gyromagnetic ratio and Gilbert 
damping constant. $\vec H_{\rm eff}=\frac{2A}{\mu_0M_{\rm s}} \nabla^2\vec m+
\frac{2K_{\rm u}}{\mu_0M_{\rm s}}m_z\hat z+H\hat z+\vec H_{\rm d}+\vec H_{\rm DM}+
\vec{h}$ is the effective field including the exchange field, the anisotropy 
field, the external magnetic field along $\hat z$, the demagnetizing field, the 
DMI field $\vec H_{\rm DM}$, and a temperature-induced random magnetic field of 
magnitude $h=\sqrt{2\alpha k_{\rm B} T/(M_{\rm s} \mu_0 \gamma\Delta V\Delta t)}$, 
where $\Delta V$, $\Delta t$, and $T$ are the cell volume, time step, and the 
temperature, respectively \cite{Brown,MuMax3}. 

In the absence of energy sources such as an electric current and the heat 
bath, the LLG equation describes a dissipative system whose energy can only 
decrease \cite{xrw1,xrw2}. Thus, solving the LLG equation is an efficient way 
to find the stable spin textures of Eqs. \eqref{energy1} and \eqref{energy2}.
The typical required time for a sample of $200\,$nm is about 
1 ns that can be estimated as follows. The spin wave speed is over $1000\,$m/s, 
thus two spins $200\,$nm apart can communicate with each other within a time of 
$0.2\,$ns. If two spins can relax after 5 times of information-exchange, it gives 
a relaxation time of $1\,$ns that is much longer than the individual dynamic 
time $h/(k_{\rm B}T_{\rm c})\simeq 1.6\, $ps if we use $T_{\rm c}\simeq 30\,$K.

This method is much faster and reliable than Monte Carlo simulations and direct 
optimizations of predefined spin textures \cite{Leonov2016,Buhrandt2013,DOYI2009,
han2010,Rybakov2016}. People \cite{han2010} have used predefined spin structures 
to determine whether helical states or SkX is more stable than the other. 
Such an approach requires a superb ability of guessing solutions of a complicated 
nonlinear differential equation in order to obtain correct physics, a formidable 
task. Very often, this approach ends with an inaccurate trial solution. 
The inaccuracy can easily be demonstrated by using the trial solution as the 
input of a simulator such as MuMax3 and seeing how far it ends at a steady state. 
We do not know any work used a reasonably good profile for stripe skyrmions in 
this approach. This may be why people using this approach did not carry out a 
self-consistent check because it is doomed to fail. 

Interestingly and unexpectedly, for a magnetic film described by energy 
\eqref{energy1} or \eqref{energy2} with 5 parameters, stable/metastable 
structures are determined by only $\kappa=\frac{\pi^2 D^2}{16AK}$ and 
$\kappa'=\frac{\pi^2D^2}{8\mu_0AM_sH}$. Let us use the bulk DMI as an example 
to prove this assertion, 
\begin{equation}
	\begin{aligned}
E=&\int{\left\{A{(\nabla\hat{m})}^2+D\left[\hat{m}\cdot(\nabla\times\hat{m})\right]
-K\left(m_z\right)^2-\mu_0M_sHm_z\right\}\ dV}\\
&=\frac{\pi^2D^2}{16A}\int \biggl\{ \left(\frac{4A}{\pi D}\nabla\hat{m}\right)^2+
\frac{4}{\pi}\left[\hat{m}\cdot\left(\frac{4A}{\pi D}\nabla\times\hat{m}\right)\right] \\ 
&\quad -\frac{1}{\kappa}\left(m_z\right)^2-\frac{2}{\kappa'}m_z \biggr\}\,dV.
	\end{aligned}
\end{equation}

Replace $\vec{x}$ by dimensionless variable $\vec{x}'=\frac{\pi D}{4A}\vec{x}$ and then 
relabel $\vec{x}'$ back to $\vec{x}$, the steady spin structures satisfy equation 
\begin{equation}
\nabla^2\hat{m}+\frac{4}{\pi}\left(\nabla\times\hat{m}\right)+\frac{1}{\kappa} m_z\hat{z}
+\frac{1}{\kappa'}\hat{z}=0. 
	\label{sseq1}
\end{equation} 
$A/D$ is a natural length of the system. 
Similarly, for the case of interfacial DMI, the equation is 
\begin{equation}
\nabla^2\hat{m}+\frac{4}{\pi}\left[(\nabla \cdot \hat{m})\hat{z}-\nabla 
m_z\right]+\frac{1}{\kappa}m_z\hat{z}+\frac{1}{\kappa'}\hat{z}=0. 
	\label{sseq2}
\end{equation} 
Both equations depend only on $\kappa=\frac{\pi^2 D^2}{16AK}$ and $\kappa'=\frac{\pi^2D^2}{8\mu_0AM_sH}$. 
To shed some light on the meaning of these two parameters, we note that exchange interaction ($A$), 
anisotropy ($K$) and external magnetic field ($H$) prefer spins aligning along the same direction 
while DMI ($D$) prefers spin curling. $K$ and $H$ are two different anisotropy parameters. 
We can further see the similarity of $\kappa$ and $\kappa'$ by examine the equations for static 
stripe skyrmions whose profile is $\hat{m}=(\cos \gamma\sin\Theta, \sin\gamma\sin\Theta,\cos\Theta)$ 
and $\gamma=0$ for interfacial DMI and $\gamma=\pi/2$ for bulk DMI \cite{xrw,haitao2,haitao}. 
From Eqs. \eqref{sseq1} and \eqref{sseq2}, equation for $\Theta$ is $\frac{\partial^2 \Theta}
{\partial x^2}{\color{blue}-}\frac{1}{2\kappa}\sin 2 \Theta {\color{blue}-} \frac{1}{\kappa'}\sin \Theta=0$ with boundary 
conditions of $\Theta(x=0)=0$ and $\Theta(x=2L)=2\pi$, here $2L$ being the stripe width \cite{haitao2}.
In the absence of a magnetic field, it is known that $\kappa=1$ separates an isolated circular skyrmion 
from condensed stripe skyrmions \cite{xrw,haitao2}. We expect then that, in the absence of anisotropic 
constant $K=0$, $\kappa'=4$ separates an isolated circular skyrmion from condensed stripe skyrmions. 
Thus both parameters describe the competition between collinear order and curling (spiral) order. 
With that being said, two anisotropies are also fundamentally different. $H$ is obviously 
unidirectional, i.e. spins parallel and anti-parallel to the field are not same, in contrast 
to the bidirectional nature of $K$. We will also see below how $H$ can make a SkX to be 
the lowest energy state while $K$ does not have such an effect. 

\section{Results and discussions}

Several sets of very different model parameters relevant to various bulk chiral magnets
are used to illustrate our main findings. These sets are listed in Table~\ref{table} 
as samples A, B, and C. Since DMI in films is mostly interfacial, we replace the bulk DMI 
by the interfacial DMI in most of our studies. However, the results are essentially the 
same no matter which DMI is used. The verification of this assertion is given in section 3.7.
\begin{table*}
	\small
	\caption{\ Three sets of parameters}
	\label{table}
	\begin{tabular*}{\textwidth}{@{\extracolsep{\fill}}llllllll}
		\hline
		Sample &  $A\,$(pJ/m) & $D\,\rm (mJ/m^2)$  & $K_u\,\rm (mJ/m^3)$ & 
		$M_{\rm s}\,\rm(MA/m)$ & $H\,$(T)& DMI type & Materials\\ \hline A & 
		0.4 & 0.33 &0.036 &0.15& variable  & itf. \&  bulk & $\rm MnSi$ 
		\cite{Fert,Karhu}      \\  \hline B & 
		0.39 & 0.0544 &0.00103 &0.03& 0.014  & bulk & $\rm Fe_{0.5}Co_{0.5}Si$ 
		\cite{FeCoSi}      \\ \hline
		C & 5  & 2 & 0.063&0.15& 0.3 & bulk &$\rm Co_8Mn_8Zn_4$ \cite{CoMnZn}
		\\ \hline
	\end{tabular*}
\end{table*}

Stripe skyrmions exist in all three samples with $\kappa>1$ \cite{xrw,haitao2}. 
The typical spin precession time is order of THz, and spin relaxation time across 
the sample is order of nano-seconds as demonstrated in both $E(t)$ and $Q(t)$. 
It should be pointed out that all results reported here are similar although very 
different sets of parameters are used. These parameters mimic very different materials 
whose Curie temperature differ by more than 10 times. Periodic boundary conditions are 
used to eliminate boundary effects and the MuMax3 package \cite{MuMax3} is employed to 
numerically solve the LLG equation for films of $200\,$nm$\times 200\,$nm$\times 8\,$nm. 
The static magnetic interactions are fully included in all of our simulations. The mesh 
size is of $1\,{\rm nm}\times1\,{\rm nm}\times 1\,{\rm nm}$ unless otherwise stated. 
The number of stable states and their structures should not depend on the Gilbert 
damping constant. We use a large $\alpha=0.25$ to speed up our simulations. \\ \par

\subsection{Metastable spin structures and density dependence of skyrmion morphology}

\begin{figure*}
	\centering
	\includegraphics[width=17cm]{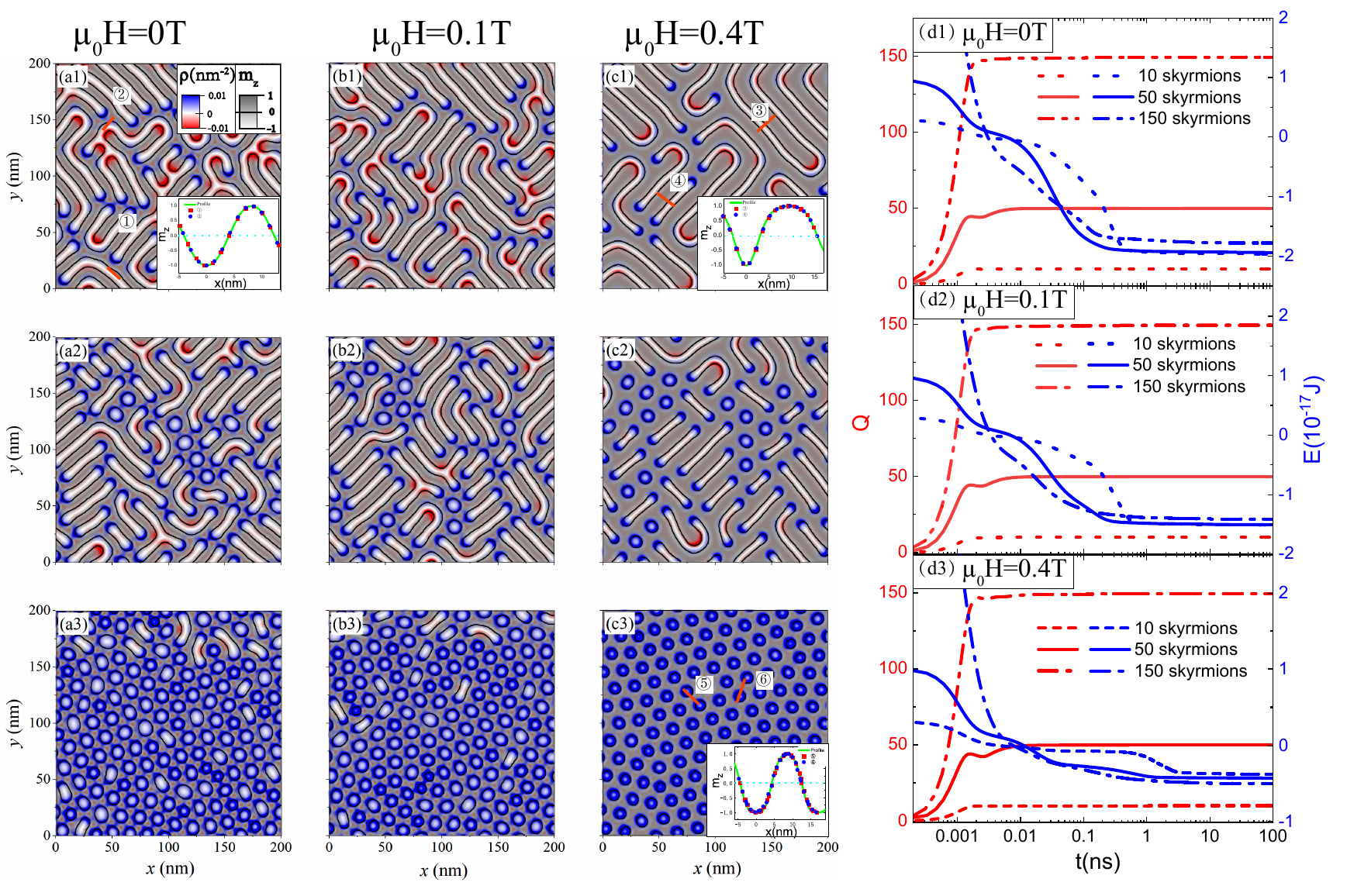}\\
	\caption{Metastable structures of 10 (a1-c1), 50 (a2-c2) and 150 (a3-c3) skyrmions 
		in a $200 {\rm nm}\times 200{\rm nm}\times 8\,$nm film at zero temperature for 
		$\mu_0H=0\,$T (a1-a3), $0.1\,$T (b1-b3) and $0.4\,$T (c1-c3), respectively, from 
		MuMax3 \cite{MuMax3}. The insets of (a1), (c1), (c3) are the spin profiles along 
		the red lines. The enlarged figures and descriptions are in Fig. 2. 
		(d1-d3) $Q$ (the left $y$-axis and the red curves) and $E$ (the right $y$-axis 
		and the blue curves) of 10 (the dash lines), 50 (the solid lines), and 150 
		(the dot-dash lines) skyrmions as functions of the time (in the logarithmic 
		scale) for $\mu_0H=0\,$T (d1), $0.1\,$T (d2) and $0.4\,$T (d3). 
		The initial configurations are 10, 50 and 150 nucleation domains of zero 
		skyrmion number and of $5\,$nm in diameter each arranged in a square lattice.}
	\label{fig1}
\end{figure*}

Figure 1 shows typical stable structures of 10, 50, and 150 skyrmions in a $200\,$nm
$\times 200\,$nm$\times 8\,$nm film under various perpendicular magnetic fields of 
$\mu_0H=0,\ 0.1$, and $0.4\,$T for sample A with $\kappa=\frac{\pi^2 D^2}{16AK}=8.3$. 
They are steady-state solutions of the LLG equation at zero temperature with 
initial configurations of 10, 50, and 150 nucleation domains of $m_z=-1$ in the 
background of $m_z=1$. Each domain is $5\,$nm in diameter, and domains are initially 
arranged in a square lattice. Figs. \ref{fig1}(a1-a3) are zero field structures 
of 10 (a1), 50 (a2), and 150 (a3) skyrmions characterized by skyrmion number $Q$ 
defined as $Q=\int \rho\,{\rm d}x{\rm d}y$, here $\rho=\frac{1}{4\pi}\hat{m}\cdot
(\partial_x \hat{m}\times \partial_y \hat{m})$ is the skyrmion charge density. 
The colour encodes the skyrmion charge distribution. Each small domain of initial 
zero skyrmion number becomes a skyrmion of $Q=1$ after a few picoseconds. 
Interestingly, both positive and negative charges appear in a stripe skyrmion while 
only positive charges exist in a circular skyrmion. Figures \ref{fig1}(d1-d3) show  
how total $Q$ (the left $y$-axis and the red curves) and energy $E$ (the right 
$y$-axis and the blue curves) change with time. $Q$ reaches its final stable 
values within picoseconds while $E$ decreases to it minimal values in nanoseconds. 
The negative skyrmion formation energy explains well the stripe skyrmion 
morphology that tries to fill up the whole film in order to lower its energy, 
in contrast to circular skyrmions for positive formation energy \cite{Xiansi}. 
Unexpectedly, the film can host an arbitrary number of skyrmions up to a 
large value (see Fig. 4 below). At a low skyrmion number of $Q=10$, the film 
is in a helical state consisting of ramified stripe skyrmions of well-defined 
width of $8.1\,$nm. Spin profile of stripes are well characterized by width 
$L$ and wall thickness $w$. Stripe width depends on materials parameters as 
$L=a(\kappa)A/D$, where $a=2\pi$ for $\kappa>>1$ \cite{xrw}. 
The film with $Q=50$ is in a helical state consisting of rectangular stripe 
skyrmions while it is a SkX of triangular lattice with $Q=150$ skyrmions. 
Skyrmion density for $Q=150$ is so high that the 
distance between two neighbouring skyrmions is comparable to the stripe width, 
and skyrmion-skyrmion repulsion compress each skyrmion into a circular object. 
Skyrmions at high skyrmion density prefer a triangular lattice as shown in 
Fig. \ref{fig1}(a3) instead of the initial square lattice.

Figures \ref{fig1}(b1-b3 and c1-c3) plot the metastable structures under 
fields $\mu_0H=0.1\,$T (b1-b3) and $\mu_0H=0.4\,$T (c1-c3) with the same 
initial states as those in (a1-a3). Stripes of $m_z>0$ (the grey regions) 
parallel to $H$ expand while the white regions of $m_z<0$, anti-parallel 
to $H$, shrink. This is showed in Figs. \ref{fig1} (b1, b2, c1, and c2). 
Moreover, the amount of increase and decrease of white and grey stripes are not 
symmetric such that skyrmion-skyrmion repulsion is enhanced by the field, and 
SkXs tend to occur at lower skyrmion density. This trend can be clearly seen 
in Figs. \ref{fig1}(c2) and (c3). Figures \ref{fig1}(d2) and (d3) show similar 
behaviour for $\mu_0H=0.1\,$T (d2) and $0.4\,$T (d3) as their counterparts 
(d1) at zero field: Skyrmion number $Q$ grows to its final values rapidly in 
picoseconds and $E$ monotonically decreases to its minimum in nanoseconds. 
Figure \ref{fig1} shows unexpectedly that the nature shape of skyrmions are 
various types of stripes when $\kappa>1$ and at low skyrmion density, 
in contrast to the current belief that all skyrmions are circular. 

\begin{figure*}[htbp]
	\centering
	\includegraphics[width=17cm]{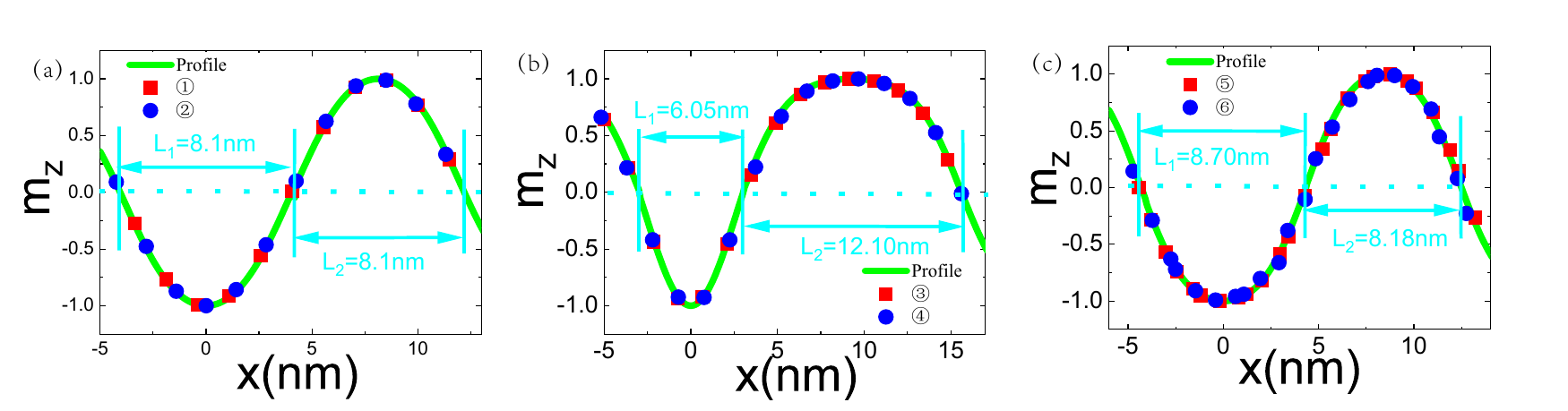}
	\caption{The red and blue points in (a), (b) and (c) are the $z$-component of 
		magnetization along the red lines in Figs. 1(a1), (c1) and (c3). 
		The green lines are the fit to our approximate profile. The cyan arrows 
		indicate the widths of stripes of $m_z<0$ or $m_z>0$ and skyrmion diameter.}
	\label{fig2}
\end{figure*}

Spin profiles of stripe skyrmions in the presence of a magnetic field are 
described by \cite{xrw,haitao2}  
$\Theta (x)=2\arctan\left[\frac{\sinh(L_1/2w_1)}{\sinh(|x|/w_1)}\right]$ 
($m_z<0$) and $\Theta (x)=2\arctan\left[\frac{\sinh(|x|/w_2)}{\sinh
	(L_2/2w_2)}\right]$ ($m_z>0$), respectively, with $|x| \leq L_i/2$. 
$\Theta$ is the polar angle of the magnetization at position $x$ and 
$x=0$ is the center of a stripe where $m_z=-1,1$ respectively. 
$L_i$ and $w_i$ ($i=1, 2$) are the width and wall thickness of stripes. 
The spin profile of skyrmions in a SkX can be well described by 
$\Theta (r)=2\arctan\left\{\frac{\tan\left[\frac{\pi}{2}\cos(\frac{\pi R}{L})
	\right]}{\tan\left[\frac{\pi}{2}\cos(\frac{\pi r}{L})\right]}\right\}$ 
along a line connecting the centers of two neighbouring skyrmions, where $r$ 
labels the points on the line with $r=0$ being the center of a chosen skyrmion. 
$R$ and $L$ are respectively the skyrmion size (the radius of $m_z=0$ contour) 
and the lattice constant. $L$ relates to skyrmion number density $n$ of 
a SkX in a triangular lattice as $L\approx 1.07/\sqrt{n}$ \cite{haitao2}.  
Figure \ref{fig2}(a) demonstrates the excellence of this approximate spin 
profile for those stripe skyrmions labelled by \textcircled{n} in Figs. 
\ref{fig1}(a3-c3) (with model parameters of sample A in Tab.~\ref{table}). 
The $y-$axis is $m_z$ and $x=0$ is the stripe center where $m_z=-1$. 
Symbols are numerical data and the solid curve is the fit of $\cos\Theta (x)=-
\frac{\sinh^2(L_1/2w_1)-\sinh^2(x/w_1)}{\sinh^2(L_1/2w_1)+\sinh^2(x/w_1)}$ for 
$-L_1/2<x<L_1/2$ and $\cos\Theta (x)=\frac{\sinh^2(L_2/2w_2)-\sinh^2\left[(x-L_1/2
	-L_2/2)/w_2\right]}{\sinh^2(L_2/2w_2)+\sinh^2\left[(x-L_1/2-L_2/2)/w_2\right]}$ 
for $L_1/2<x<L_1/2+L_2$ with $L_1=L_2=8.1\,$nm and $w_1=w_2=2.7\,$nm. 
All data from different stripes fall on the same curve and demonstrate that 
stripes, building blocks of the structure, are identical. Figure \ref{fig2}(b) 
shows the nice fit of numerical data of all stripes under a magnetic field of 
$0.4\,$T to the approximate spin profile with $L_1=6.05\,$nm, $L_2=12.10\,$nm, 
and $w_1=w_2=2.63\,$nm. Figure \ref{fig1}(c) shows that spins along a line 
connecting the centers of two neighbouring skyrmions fall on our proposed 
profile with $L=16.88\,$nm, and $R=4.35\,$nm. The spin profile has 
been used to find stripe width formula $L=f(\kappa) A/D$ \cite{xrw}. 
The value of $f(\kappa)$ is $2\pi$ when $\kappa \gg 1$. 

\par 

\subsection{Stable/metastable spin structures defined by  
	$\kappa$  and $\kappa'=\frac{\pi^2D^2}{8\mu_0AM_sH}$} 
\begin{figure*}[htbp]
	\centering
	\includegraphics[width=17cm]{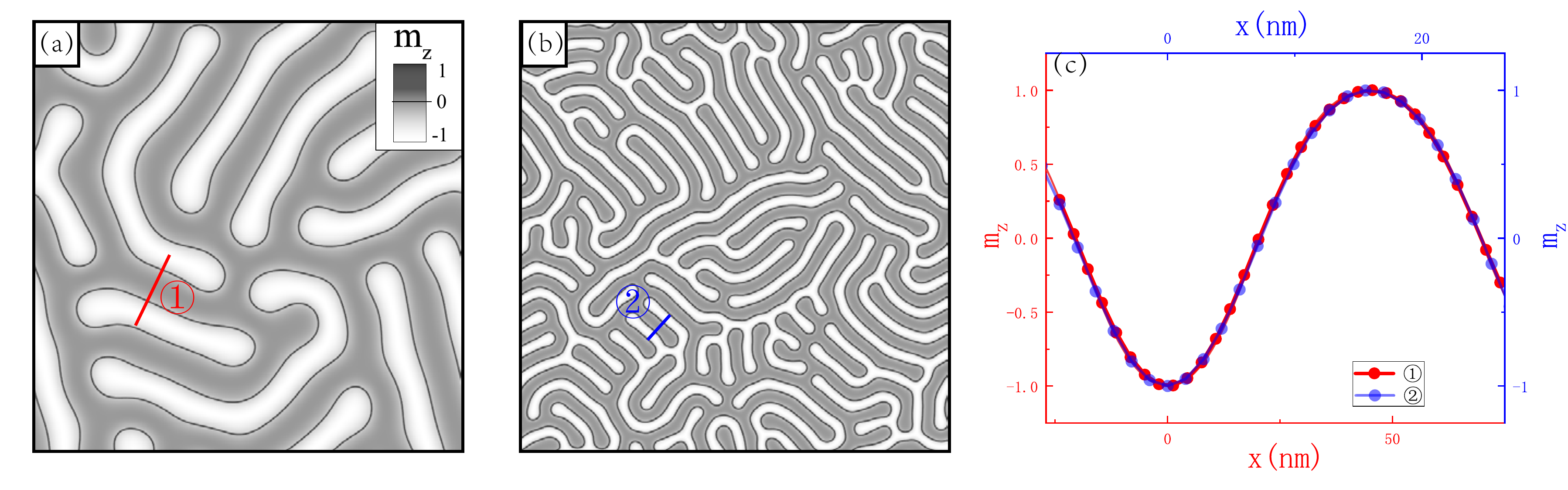}
\caption{Metastable structures of 10 skyrmions in a $600\,$nm$\times600\,$nm$\times1\,$nm 
film of $\kappa=10$ and $\kappa'=22$ for samples B (a) and C (b) in Table\ref*{table}. 
(c) Spin profiles along the red and blue lines in (a) and (b). The x-axis is scaled 
by $A/D$. The red curve and the bottom x-axis are for sample B while the blue curve and 
the top x-axis are for sample C.}
	\label{fig3}
\end{figure*}

We have analytically proved in the previous section that stable/metastable 
structures are fully determined by $\kappa$ and $\kappa'$ out of five  
model parameters, and $A/D$ defines the length scale. In another word, 
the spin structures of the same $\kappa$ and $\kappa'$ are scale invariant. 
To verify the results, we use samples B and C listed in Table~\ref{table} to 
simulate metastable structures of 10 skyrmions in a film of $600\,$nm$\times 
600\,$nm$\times 1\,$nm as what was done in Fig. 1. Both sample B and sample C 
have the same $\kappa=10$ and $\kappa'=22$ although their $A$ and $D$ differ 
by more than ten times. The stable structures are similar as shown in 
Figs.~\ref{fig3}(a) for sample B and \ref{fig3}(b) for sample C. 
To further verify that the two structures are scaled by $A/D$, the spin profiles 
along the red and blue lines in Figs.~\ref{fig3}(a) and (b) are presented in 
Fig.~\ref{fig3}(c). When we scale the x-axis coordinate by $A/D$ for sample B 
(the red curve and bottom x-axis) and sample C (the blue curve and top x-axis), 
two curves are overlapped that verifies our theoretical predictions.

\par 

\subsection{The role of magnetic field in SkX formation} 

\begin{figure*}
	\centering
	\includegraphics[width=17cm]{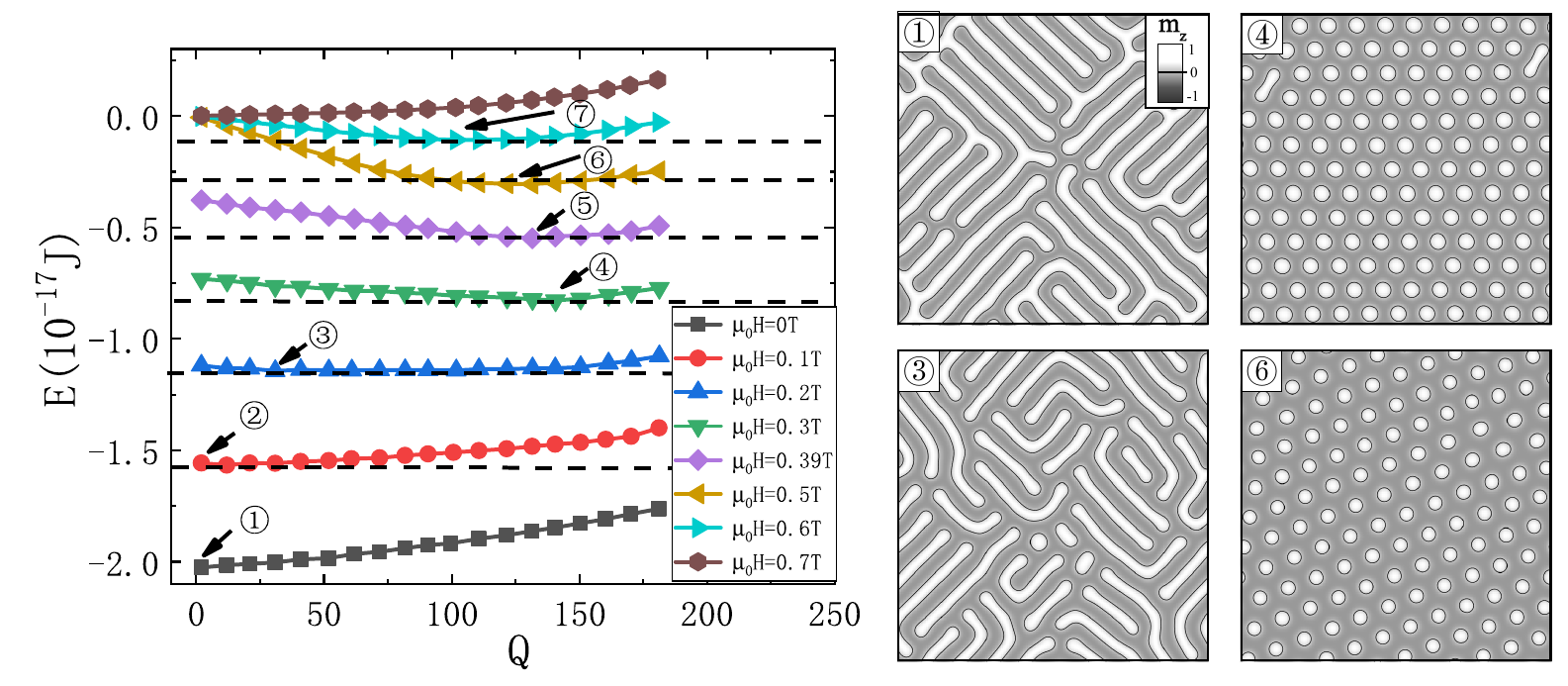}\\
	\caption{Skyrmion-number dependence of energy at various magnetic fields of 
		$\mu_0H=0\,$T (the squares and the black curve), $0.1\,$T (the circles and 
		the red curve), $0.2\,$T (the up-triangles and the blue curve), $0.3\,$T 
		(the down-triangles and the green curve), $0.39\,$T (the diamonds and the 
		violet curve), $0.5\,$T (left-triangles and the yellow curve), $0.6\,$T 
		(right-triangles and the cyan curve), and $0.7\,$T (the stars and the brown 
		curve) (the left panel). The film size is the same as in Fig. \ref{fig1}. 
		Arrows denote the $Q_{\rm m}$ for $\mu_0H=0, \ 0.1, \ 0.2,\ 0.3,\ 0.39, \ 0.5$ 
		and $0.6\,$T, respectively. Corresponding structures are shown in the right panel. 
		The structures varies from a highly ramified stripe of $Q_{\rm m}=1$ for 
		$\mu_0H=0\,$T, to the mixture of helical order and SkX of $Q_{\rm m}=31$ for 
		$\mu_0H=0.2\,$T, and to beautiful SkX of $Q_{\rm m}=141$ for $\mu_0H=0.3\,$T. 
		Equally nice SkX for $\mu_0H=0.5,\ 0.6\,$T with a smaller $Q_{\rm m}$ 
		demonstrates importance of a field in SkX formation.}
	\label{fig4}
\end{figure*}

We compute below the skyrmion-number-dependence of energy for various  
magnetic field in order to understand the role of a field in SkX formation. 
We limit this study to the $Q$'s far below its maximal value of more than 500. 
Physics around the maximal $Q$ is very interesting by itself, but not our 
concern here, and it will be investigated in the future. $Q$ nucleation domains 
of $5\,$nm in diameter each arranged in a square lattice is used as the initial 
configuration to generate a stable structure of $Q$ skyrmions with energy $E$. 
The left panel of Fig. \ref{fig4} is the $Q$-dependence of $E$ from MuMax3 
simulations for $\mu_0 H=0,$ 0.1, 0.2, 0.3, 0.39, 0.5, 0.6 and $0.7\,$T denoted 
by \textcircled{1}-\textcircled{7} . At zero field, $E$ increases monotonically 
with skyrmion number $Q$. States with few skyrmions or low skyrmion density 
are preferred. Since long stripe skyrmions have more way to deform than 
circular skyrmions, the entropy of a helical phase is larger than a SkX (see 
numerical evidences according to the approach in Ref. \cite{Bocarsly2018} below). 
Thus, helical states should always be the thermal equilibrium phase below the 
Curie temperature, and a SkX can be a metastable state at most. 
$E$ of helical states at zero field is not very sensitive to $Q$. 
Thus the thermal equilibrium helical states can have different 
number of irregular skyrmions with many different forms or morphologies. 
This understanding agrees with experimental facts of rich stripe morphologies 
\cite{roadmap}. Things are different when a magnetic field is applied. 
Firstly, $E$ of fixed $Q$ increases with $H$. Secondly, $E$ is minimal at $Q_{\rm m}$ 
for a fixed field below a critical value. $Q_{\rm m}$ first increases with $H$ up to 
an optimal field of around $\mu_0H=0.3\,$T in our case and then decreases with $H$. 
Above $\mu_0H=0.7\,$T (the brown stars), positive $E$ means that ferromagnetic state 
of $m_z=1$ has a lower energy of $E=0$. Thus, $m_z=1$ is the thermal equilibrium 
state below the Curie temperature when $\mu_0H>0.7\,$T. $Q_{\rm m}$, at the optimal 
field of $\mu_0 H=0.3\,$T, can be as large as more than 141 or a skyrmion density 
more than $3500/\,\rm\mu m^2$ at which two nearby skyrmions are in contact. 
All skyrmions are compressed into circular objects and form a SkX. 
Strictly speaking, they are not circular, as evident from our simulations. 
Our results agree qualitatively with experiments \cite{roadmap,Romming,
	Wang2020,legrand2017,Yu2011}. 

\par

\begin{figure}[htbp]
	\centering
	\includegraphics[width=\columnwidth]{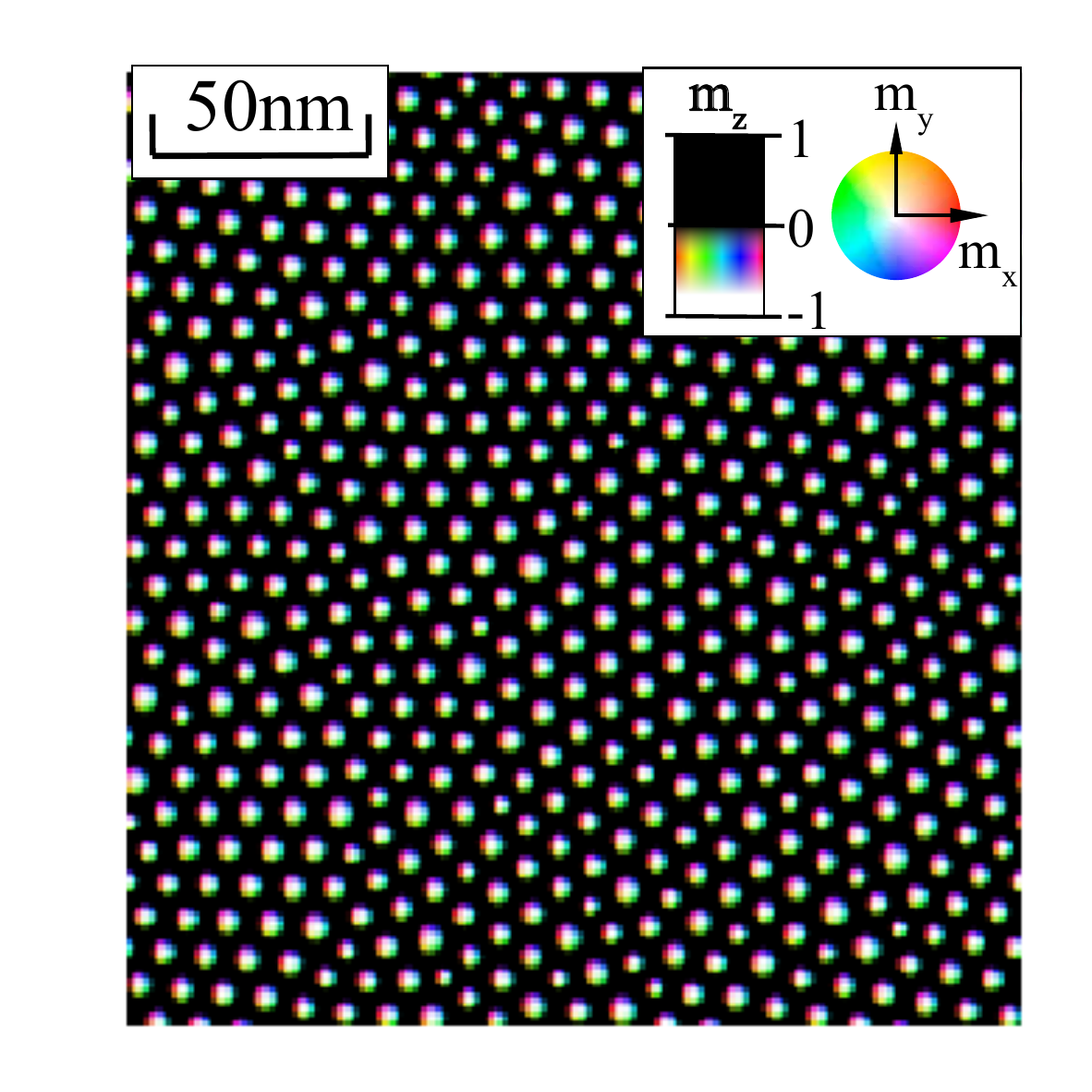}
	\caption{A SkX of 567 skyrmions when 625 nucleation domains of radius 
		$2\,$nm each are initially arranged in square lattice.}
	\label{fig5}
\end{figure}

To substantiate our claim that there is a maximal skyrmion density for a given 
film, we consider a $200\,$nm$\times 200\,$nm$\times 8\,$nm film for sample A. 
To find the approximate maximal number of skyrmions that can maintain its 
metastability, we simply add more 2nm-domains in square lattices as the 
initial configurations. The system always settles to a metastable state in 
which the number of skyrmions is the same as the number of initial 
nucleation domains as long as the the number is less than 500. 
If the number is bigger than 700, the final number of skyrmions in stable 
states would be less than the initial number of nucleation domains. 
Occasionally, we obtain metastable states containing about 700 skyrmions 
in a triangular lattice, corresponding to skyrmion density of 17,500$\,/
\rm{\mu m^2}$. Figure \ref{fig5} shows a SkX of 567 skyrmions when the 
initial configuration contains 625 nucleation domains of $2\,$nm radius 
each arranged in a square lattice. The details of how the final metastable 
states depends on the initial number of nucleation domains and their 
arrangement, as well as the boundary conditions, deserves a further study.  

\par 

\subsection{Condensed skyrmion states in various thermodynamic process}

\begin{figure*}
	\centering
	\includegraphics[width=17cm]{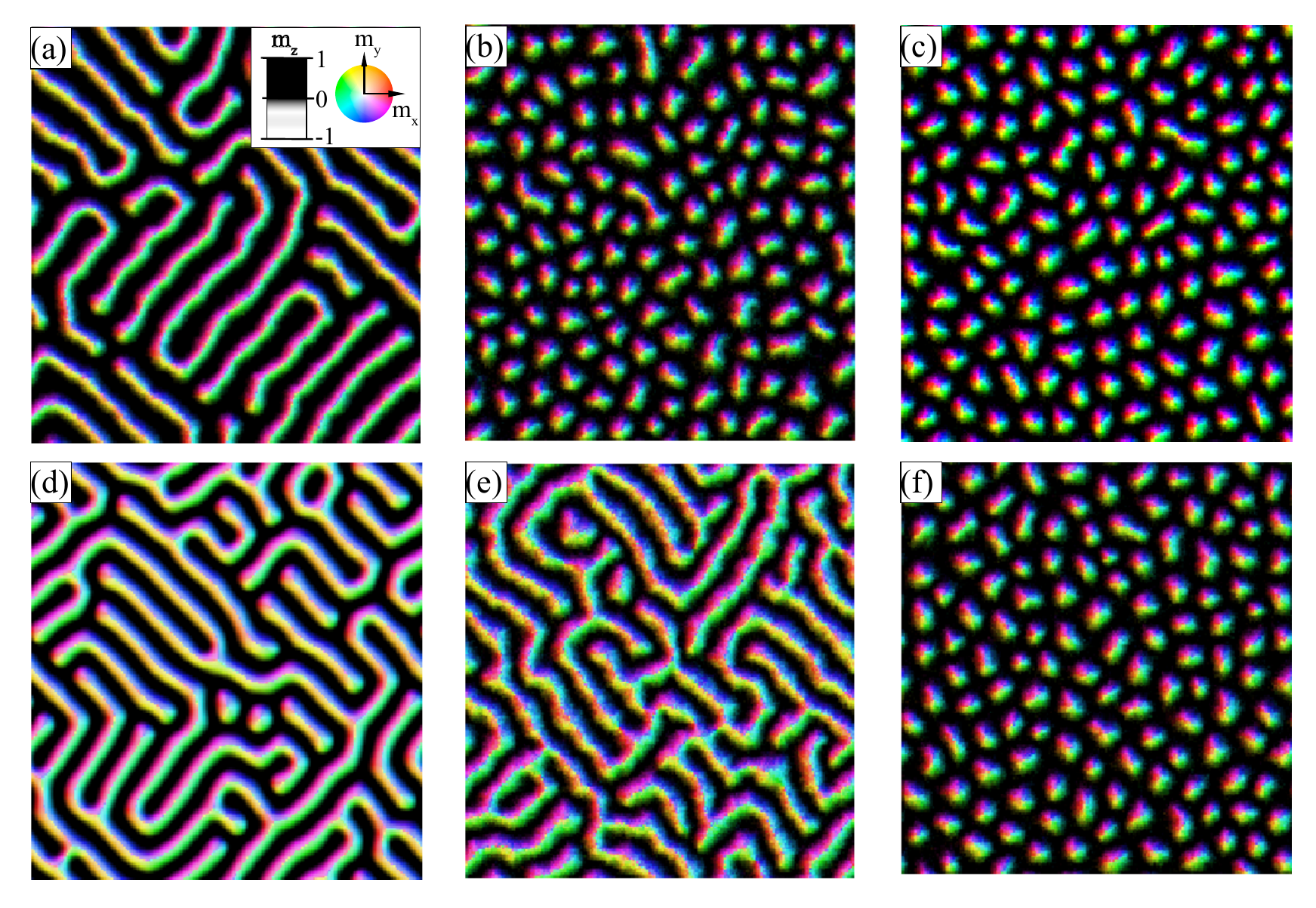}\\
	\caption{(a, b) Structures after $30\,$ns evolution at $\mu_0H=0.3\,$T, and 
		$T=20\,$K (a) and $29\,$K (b), starting from the initial configuration of 
		Fig. \ref{fig1}(c1). (c) Structures after $10$ns evolution at $\mu_0H=0.3\,$T 
		and $T=29\,$K, starting from the initial configuration of Fig. \ref{fig1}(c3).
		(d,e) Structures after $30\,$ns zero field cooing (d) and warming (e), starting 
		from the SkX in Fig. \ref{fig1}(c3). (f) A SkX of 133 skyrmions at $\mu_0H=0.3
		\,$T and $29\,$K comes from the thermally generated nucleation centers. }
	\label{fig6}
\end{figure*}

To substantiate our assertion that topology prevents a metastable helical 
state from transforming into a thermal equilibrium SkX state at $\mu_0H=0.3\,$T 
and far below the Curie temperature, we study stochastic LLG at a finite 
temperature starting from the structure of Fig. \ref{fig1}(c1) with 10 skyrmions. 
We run the MuMax3 \cite{MuMax3} at $20\,$K and $29\,$K, respectively below and 
near the Curie temperature of $T_{\rm c}=33\,$K. Figures~\ref{fig6}(a) and (b) 
show the structures after $30\,$ns evolution. Thermal fluctuations at $20\,$K are 
not strong enough to neither create enough nucleation centers nor cut a short 
stripe into two that are the process of destroying the conservation of skyrmion 
number, at least within tens of nanoseconds. The system is still in a helical state 
with $Q=11$ skyrmions after 30 nanoseconds, one more than the initial value (see 
Movie a in the Supporting Information). However, at $29\,$K, a helical state 
transforms to the thermal equilibrium SkX state with $Q=132$ skyrmions through 
cutting stripes into smaller pieces and creating nucleation centers to generate 
more skyrmions thermodynamically (see Movie b in the Supporting Information). 
For a comparison, we have also started from the SkX shown in Fig. \ref{fig1}(c3) 
with 150 skyrmions at $\mu_0H=0.3\,$T. The SkX under $29\,$K becomes another SkX 
with $Q=136$ skyrmions as shown in Fig. \ref{fig6}(c) after $10\,$ns evolution 
(see Movie c in the Supporting Information), 14 less than the starting value. 
This is expected because the average skyrmion number at the thermal equilibrium 
state should be smaller than $Q_{\rm m}=141$ according to the energy curve of 
$\mu_0H=0.3\,$T in Fig. \ref{fig4}. 

We demonstrate below that a SkX at zero field is not a thermal equilibrium state 
by showing the disappearance of a SkX in both zero field cooling and warming. 
Starting from the SkX in Fig. \ref{fig1}(c3) and gradually increasing 
(decreasing) the temperature from $0\,$K ($30\,$K) to $30\,$K ($0\,$K) at sweep 
rate of $1\,\hbox{K/ns}$ at $\mu_0H=0\,$T (see Movies d and e in the Supporting 
Information), final structures shown in Fig. \ref{fig6}(d) (zero field cooling) 
and (e) (zero field warming) are helical state consisting of stripe skyrmions. 
In contrast, field cooling at the optimal field of $\mu_0H=0.3\,$T from 
$30\,$K to $0\,$K at same sweep rate does not change the nature of the SkX. 
This is consistent with our assertion that SkXs are the thermal 
equilibrium states at $\mu_0H=0.3\,$T below the Curie temperature.

The nucleation centers can also be thermally generated near the Curie 
temperature such that skyrmions can develop from these thermally generated 
nucleation centers, rather than from artificially created nucleation domains. 
To substantiate this claim, we carried out a MuMax3 simulation at $29\,$K 
under the perpendicular field of $\mu_0H=0.3\,$T, Fig. \ref{fig6}(f) is a 
snapshot of spin structure of the thermal equilibrium state for the same 
film size and with the same model parameters as those for Figs. 6(a-e). 
A SkX with 133 skyrmions is observed. The birth of these skyrmions and how 
the SkX is formed can be seen from Movie f in the Supporting Information. 
\par

Small energy gain or loss from the transition between two states of different $Q$'s 
makes such a transformation difficult because of the conservation of skyrmion 
number under continuous spin structure deformation and entanglements among stripes. 
This study shows that, similar to liquid drop formation, new skyrmions can be 
generated only from nucleation centers or by splitting a stripe skyrmion into two. 
These process require external energy sources such as the thermal bath and result 
in topological protection and energy barrier between states of different $Q's$. 
Although the energy of $Q=141$ SkX has the lowest energy at $\mu_0H=0.3\,$T at 
zero temperature (Fig. 4), an initial state with a few stripe skyrmions would 
not resume its lowest energy state at a low temperature (Fig. \ref{fig6}(a)) 
within the simulation time. 
This demonstrates the multi-metastable states of various $Q$ and topological 
protection to prevent SkXs and helical states from relaxing to the thermal 
equilibrium phases. People have studied the thermal effects on stripes and SkXs 
\cite{Kwon2012}, but early studies did not revealed the role of $\kappa$ and 
magnetic field and cannot come up with a unified picture rich observations. 
The present new understanding can perfectly explain 
those fascinating appearance and disappearance of SkX and helical states along 
different thermodynamic paths \cite{skm-form-prm,Tokura-nm16}. For example, the 
disappearance of a SkX in zero field warming and high field warming is because 
helical state is the thermal equilibrium phase. At a high enough temperature 
below the Curie temperature, thermal fluctuations can spontaneously generate 
enough nucleation centers such that the system can change its skyrmion number 
and reach its thermal equilibrium phase of either helical state of low skyrmion 
density or SkX state of high skyrmion density. 

\subsection{$\kappa$ and SkX formation}

\begin{figure*}
	\centering
	\includegraphics[width=17cm]{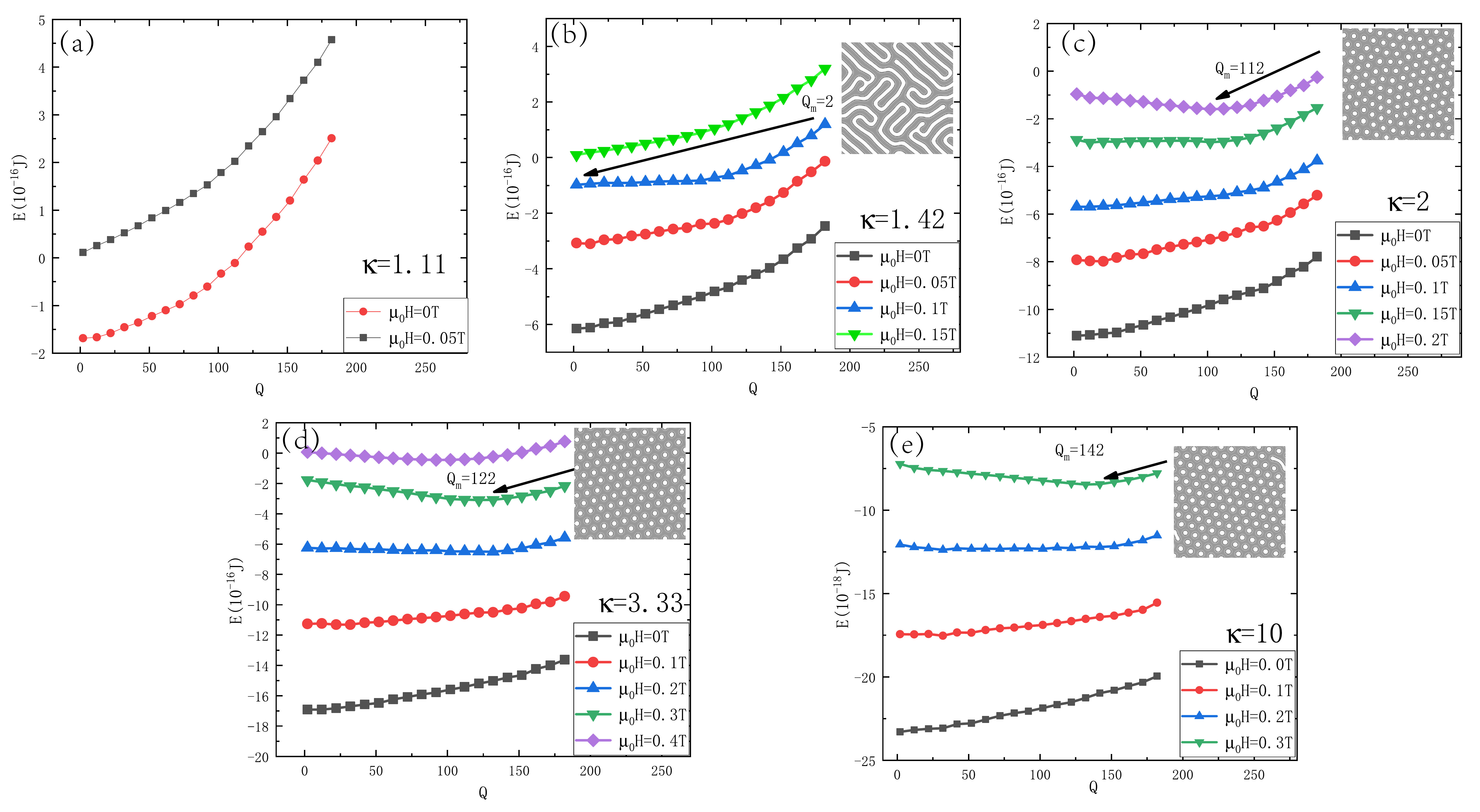}\\
	\caption{Skyrmion-number dependence of energy at various magnetic 
		fields for $\kappa=1.11$ (a), 1.42 (b), 2. (c), 3.33 (d) and 10 (e). 
		Corresponding spin textures at $Q_{\rm m}$ are shown in the insets. 
		All other model parameters are the same as those for Fig. 2.}
	\label{fig7}
\end{figure*}

All results presented so far are for relative large $\kappa>8$. It is not clear 
how the results change with other $\kappa$'s. It should be emphasized that results 
above are based on the assumptions that all model parameters do not change with 
the temperature. Obviously, this assumption does not apply to real materials. 
Exchange stiffness $A$ and magnetic anisotropy $K$ depend on temperature, and 
behave as a power law of $\vec{M}(T)$ \cite{Schlickeiser}. For example, it is known 
magnetic anisotropy in many materials can change appreciably with temperature. 
Magnetic anisotropy of $1.2\,$nm CoFeB film could reduce by 50\% as temperature 
increases from $300\,$K to $400\,$K \cite{Lee2017}. More importantly, it is 
known experimentally \cite{Oike2016,Karube2020,Crisanti2020,Birch2019} 
that SkXs are not stable at low temperature in almost all systems so far. 
Thus, a proper explanation of metastability of SkXs at low temperatures and at the 
optimal magnetic field is required. $\kappa=1$ separates condensed skyrmion states 
from isolated skyrmions \cite{haitao2}. $\kappa$, in general, decreases as the 
temperature is lowered since both $A$ and $K$ increases as the temperature decreases. 
Thus, it is interesting to find out how results in Fig. \ref{fig7} are modified when 
$\kappa=8.3$ for sample A becomes not too far from $\kappa=1$ in order to mimic the 
effect of lowering the temperature far from the Curie temperature. We repeat the 
same calculations as those in Fig. \ref{fig4} by changing crystalline magnetic 
anisotropy in our model from $0.036\,\rm MJ/m^3$ to $0.165\,\rm MJ/m^3$, $0.131\,\rm 
MJ/m^3$, $0.098\,\rm MJ/m^3$, $0.064\,\rm MJ/m^3$, and $0.030\,\rm MJ/m^3$ without 
changing other parameters. This corresponds to change $\kappa$ from 8.3 for Fig. 
\ref{fig4} to $1.11,\,1.42,\,2.,\,3.3$, and 10. The results are shown in Fig. \ref{fig7}. 
It is interesting to notice $Q_m\approx 0$ when $\kappa=1.11,\ 1.42$ such that stripy 
phase is thermal equilibrium state ($E>0$ means ferromagnetic state more stable).  
$Q_m\approx 132,\,142$ and the optimal field is $\mu_0H=0.3\,$T 
for $\kappa=3.3,\,10$, respectively. $Q_m\approx 92$ at the optimal field of 
$\mu_0H=0.15\,$T for $\kappa=2$, and the corresponding spin texture is a mixture 
of stripe skyrmions and circular skyrmions as shown in the insets of the figure. 
Our numerical simulations on two very different sets of model parameters suggest 
that $Q_m$ will not be large enough to support an thermal equilibrium SkX state 
whenever $\kappa <1.6$. 

\begin{figure*}[htbp]
	\centering
	\includegraphics[width=17cm]{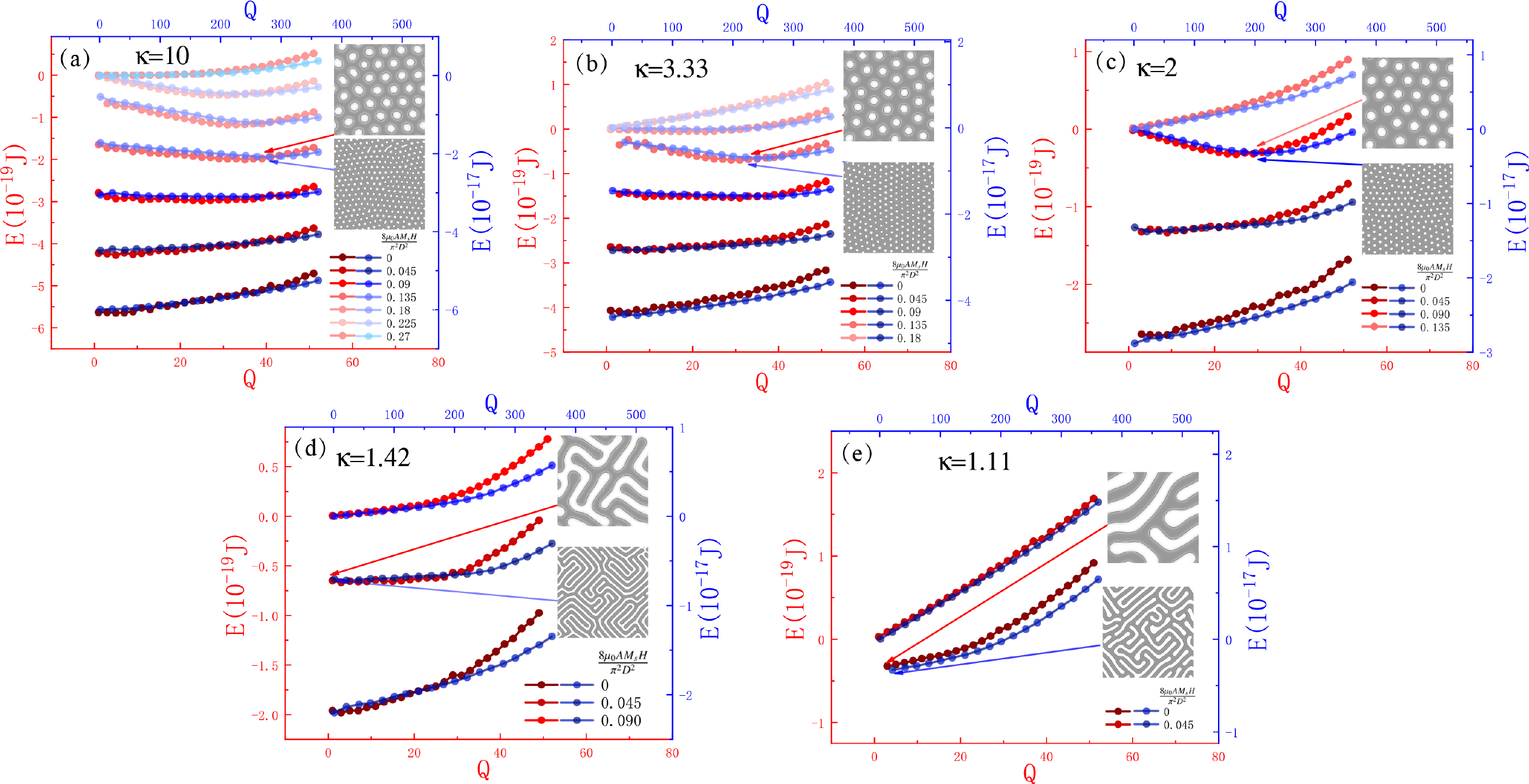}
	\caption{$E(Q)$ for sample B (red curves and axes) and sample C (blue curves 
		and axes) for various $\mu_0H$ (given by $\frac{8\mu_0AM_sH}{\pi^2D^2}$) and 
		for $\kappa=10$ (a), 3.33 (b), 2. (c), 1.45 (d), and 1.11 (e). 
		$A$, $D$ and $M_{\rm s}$ are fixed as given in Table~\ref*{table}. 
		Corresponding spin textures at $Q_{\rm m}$ are shown in the insets.
	}
	\label{fig8}
\end{figure*}

We see that $Q_m$ is not large enough to form a SkX even at the optimal field 
when $\kappa <1.6$. To see that this is in general true, we vary $\kappa$ by 
increasing the magneto-crystalline anisotropy for both samples B and C in Table
\ref{table} such that $\kappa$ changes from 10 to 3.33, 2., 1.42 and 1.11.
The film size of samples B and C is $600\,$nm$\times 600\,$nm$\times 1\,$nm.
Figure~\ref{fig8} shows $E(Q)$ for various $\mu_0H$ and for $\kappa=10$ (a),
3.33 (b), 2. (c), 1.42 (d), 1.11 (e). The red curves are for sample B while 
the blue ones are for sample C. Clearly, $Q_m$ is small and the helical state 
has the lowest energy for both sets of parameters differing by more than ten 
times whenever $\kappa<1.6$.
\par 

In summary, general decreases of $\kappa$ may explain why 
SkXs in most systems become metastable at low temperatures. 
Material parameters vary also with film thickness. This may also explain why 
the SkX formation and SkX stability are very sensitive to the film thickness. 
This conjecture needs more detail studies.

\begin{figure*}[htbp]
	\centering
	\includegraphics[width=17cm]{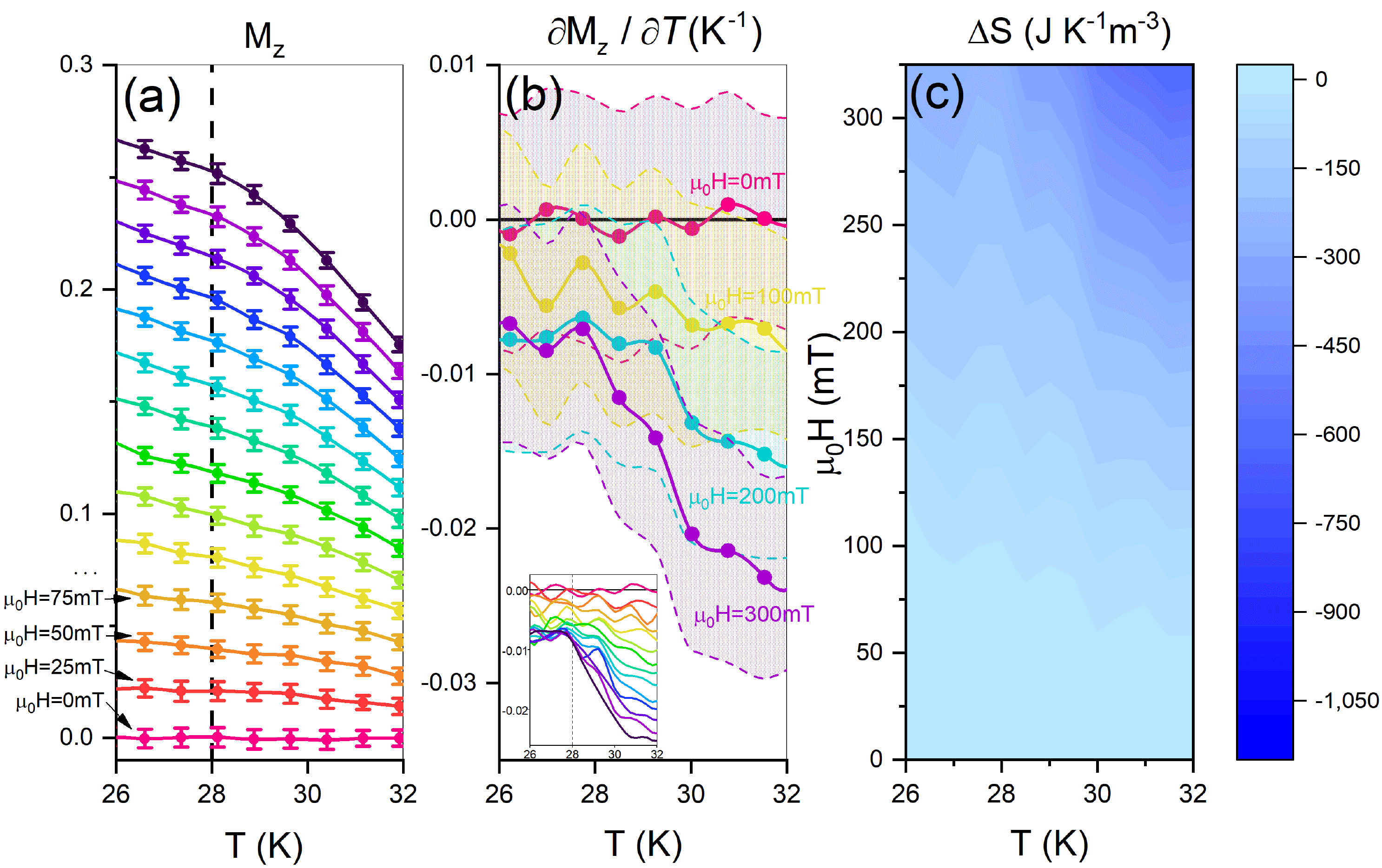}
	\caption{(a) Temperature dependence of time-averaged $z$-component of 
magnetization $M_z$ at fixed magnetic fields, ranging from $0\,$mT to 
$300\,$mT with a field interval of $25\,$mT. Every data point is 
2ns-averaged $z$-component of magnetization in the thermal equilibrium 
state at a fixed temperature and a fixed magnetic field. Error bars are the 
standard deviations of magnetization over the time periods. 
		(b) Solid curves are temperature derivatives of magnetization $\partial M_z/
		\partial T$ at $\mu_0H=0,\,100,\,200,\rm\,and\,300\,mT$. Dash curves bound 
		1-sigma regions. Inset shows $\partial M_z/\partial T$ for all field 
		strength in (a). (c)  Density plot of entropy change $\Delta S$ from a  
		stripy state at zero field to state at a finite field at fixed temperatures 
		in the $T$-$H$ plane. Clearly, a SkX at $T=28\,$K and $\mu_0H=300\,$mT has 
		lower entropy than that of a stripy state at $T=28\,$K and $\mu_0H=0\,$mT.}
	\label{fig9}
\end{figure*}\par 

\subsection{Entropy difference of SkXs and stripy states at a given temperature}
In an isothermal process, the Maxwell relation 
$(\partial S/\partial H)_T=(\partial M/\partial T)_H$ provides a way of 
calculating the entropy difference at two different fields for a given 
temperature from the temperature dependence of magnetization \cite{Bocarsly2018},
\begin{equation}
	\Delta S(T,H)\equiv S(T,H)-S(T,0)=\int_0^H(\frac{\partial M}
	{\partial T})_{H'}dH'.
	\label{DeltaS}
\end{equation} 
$\Delta S(T,H_{\rm m})$ measures the entropy change from a stripy state at zero 
field to a SkX at optimum field $H_{\rm m}$ when $T\simeq T_c$. Differ from 
experiments where the noises are bigger than the signals \cite{Bocarsly2018}, 
simulations do not suffer from this problem such that one can reliably 
compute the difference between a SkX entropy and a stripy-state entropy at a 
fixed temperature. We carried out MuMax3 simulations for $M(T,H)$ of sample 
A around $T=28\,$K for various $\mu_0H$ ranging form $0\,$T to $0.3\,$T. 
Sample size is $\rm 200\,nm\times 200\,nm\times 8\,nm$ as before. 
Figures \ref{fig9}(a-b) are $M(T,H)$ (a) and $(\partial M/\partial T)_H<0$ 
(b) for various $H$. $M(T<T_c,0)\simeq 0$ as shown in Fig. \ref{fig9}(a) 
because of our helimagnet model. This is why $(\partial M/\partial T)_{H=0}$ 
fluctuates around zero as shown by the red dots in Fig. \ref{fig9}(b) and 
the one-$\sigma$ region is bounded by two red dash lines.
One can confidently conclude that $(\partial M/\partial T)_H<0$ for $T>0.1T$ 
(the yellow dots and curves) at which a stripy state has the lowest energy 
[see Fig. \ref{fig4}(a)] while one may not be so sure of the sign of 
$(\partial M/\partial T)_H$ in experiments. The simulation results clearly 
support the claim that SkX entropy is smaller than a stripy-phase entropy.  

\subsection{Insensitivity of spin structures to DMI-types} 

There are two types of Dzyaloshinskii–Moriya interaction (DMI), namely 
interfacial DMI and bulk DMI. We have mainly presented results for the 
interfacial DMI so far. Although the spin orientations in skyrmion walls 
depends on the type of DMIs, the location of energy minimum in energy-$Q$ 
curves for a fixed $\mu_0H$, as well as the thermodynamic properties of 
the system, are not sensitive to DMI type. When the interfacial DMI 
is replaced by the bulk DMI for those sets of parameters used above,  
the results are essential the same except that the N\'{e}el-type stripes 
(skyrmions) for the interfacial DMI change to the Bloch-type stripes 
(skyrmions). 

\begin{figure*}[htbp]
	\centering
	\includegraphics[width=17cm]{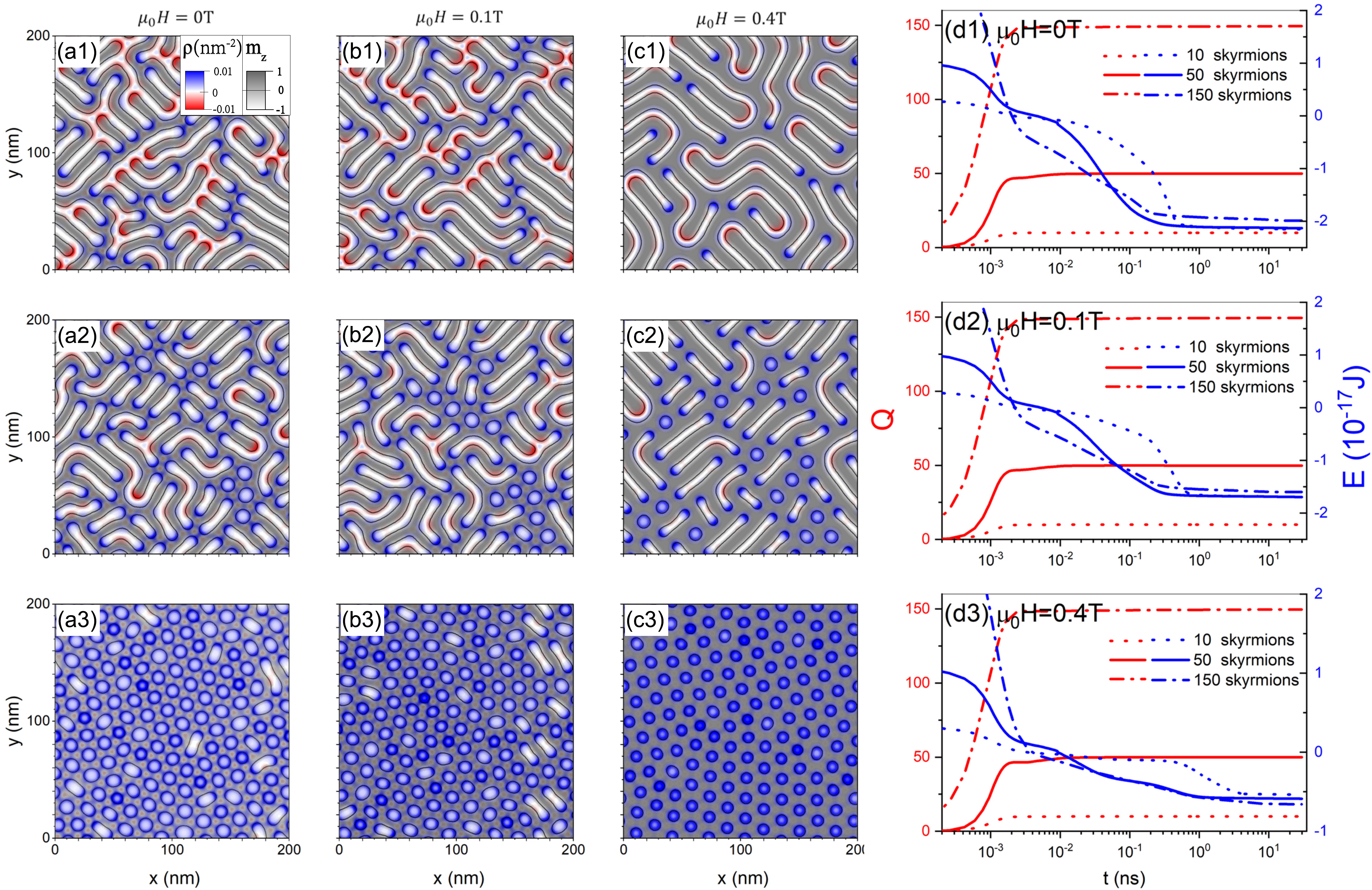}
	\caption{Metastable structures of 10 (a1-c1), 50 (a2-c2) and 150 (a3-c3) 
		skyrmions in a $200 {\rm nm}\times 200{\rm nm}\times 8\,$nm film with the bulk 
		DMI at zero temperature for $\mu_0H=0\,$T (a1-a3), $0.1\,$T (b1-b3) and $0.4\,$T 
		(c1-c3), respectively. (d1-d3) show how $Q$ (the left $y$-axis and the red curves) 
		and $E$ (the right $y$-axis and the blue curves) of 10 (the dash lines), 50 (the 
		solid lines), and 150 (the dot-dash lines) skyrmions vary with time (in the 
		logarithmic scale) for $\mu_0H=0\,$T (d1), $0.1\,$T (d2) and $0.4\,$T (d3). 
		The insets of (a1), (c1), (c3) show the spin profiles of stripe and circular skyrmions. 
		The initial configurations are 10, 50 and 150 nucleation domains arranged in a square 
		lattice. Initially, each domain is $5\,$nm in diameter and has zero skyrmion number.}
	\label{fig10}
\end{figure*}

\begin{figure*}[htbp]
	\centering
	\includegraphics[width=17cm]{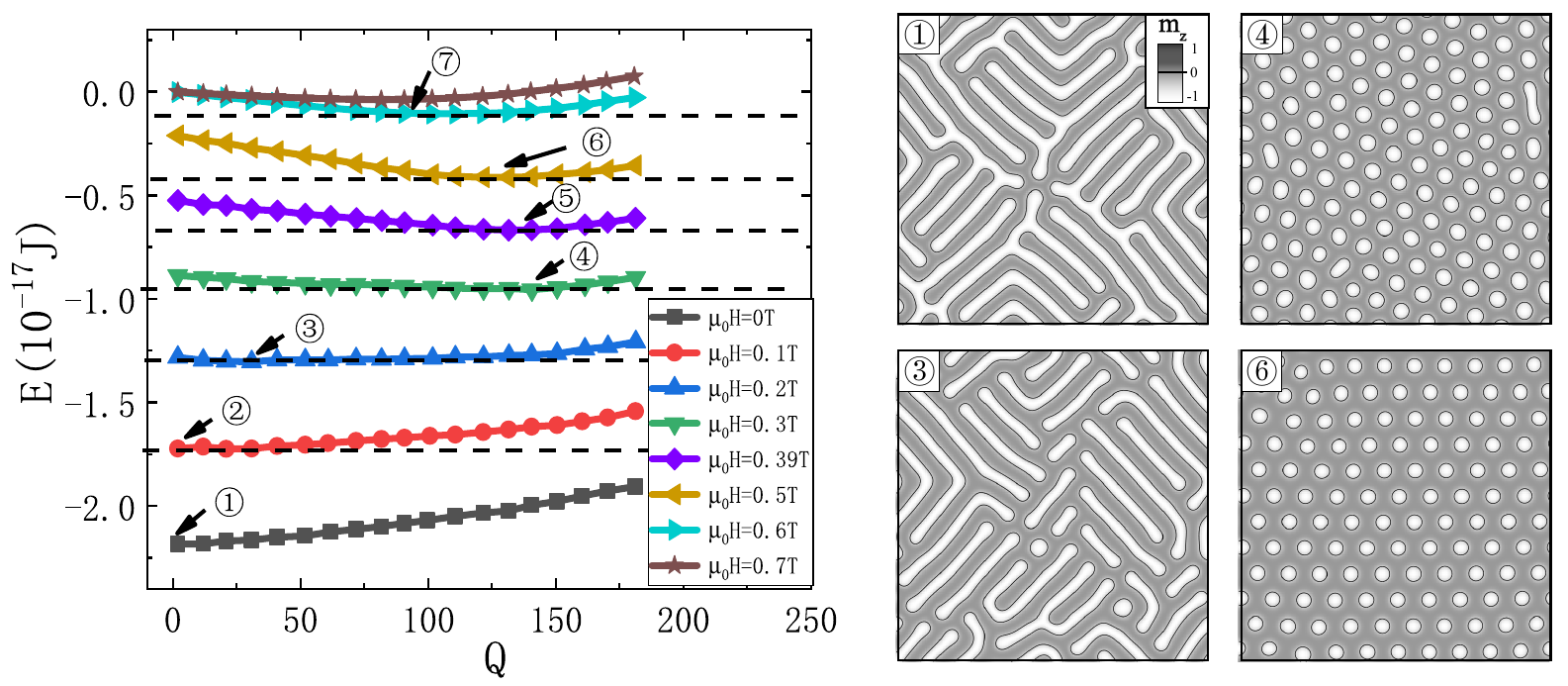}
	\caption{Skyrmion-number dependence of energy for various magnetic fields of 
		$\mu_0H=0\,$T (the squares and the black curve), $0.1\,$T (the circles and 
		the red curve), $0.2\,$T (the up-triangles and the blue curve), $0.3\,$T 
		(the down-triangles and the green curve), $0.39\,$T (the diamonds and the 
		violet curve), $0.5\,$T (the left-triangles and the yellow curve), $0.6\,$T 
		(the right-triangles and the cyan curve), and $0.7\,$T (the stars and the brown 
		curve) (the left panel). The film size is the same as that in Fig. \ref{fig10}. 
		Arrows denote the $Q_{\rm m}$ for $\mu_0H=0, \ 0.1,
		\ 0.2,\ 0.3,\ 0.39, \ 0.5$ and $0.6\,$T, respectively. 
		Corresponding structures are shown in the right panel. 
		The structures varies from highly ramified stripes of $Q_{\rm m}=1$ for 
		$\mu_0H=0\,$T, to the mixture of helical order and SkX of $Q_{\rm m}=31$ for 
		$\mu_0H=0.2\,$T, and to a beautiful SkX of $Q_{\rm m}=141$ for $\mu_0H=0.3\,$T. 
		Equally nice SkX for $\mu_0H=0.5,\ 0.6\,$T with a smaller $Q_{\rm m}$ 
		demonstrates importance of a field in SkX formation.}
	\label{fig11}
\end{figure*}

\begin{figure*}[htbp]
	\centering
	\includegraphics[width=17cm]{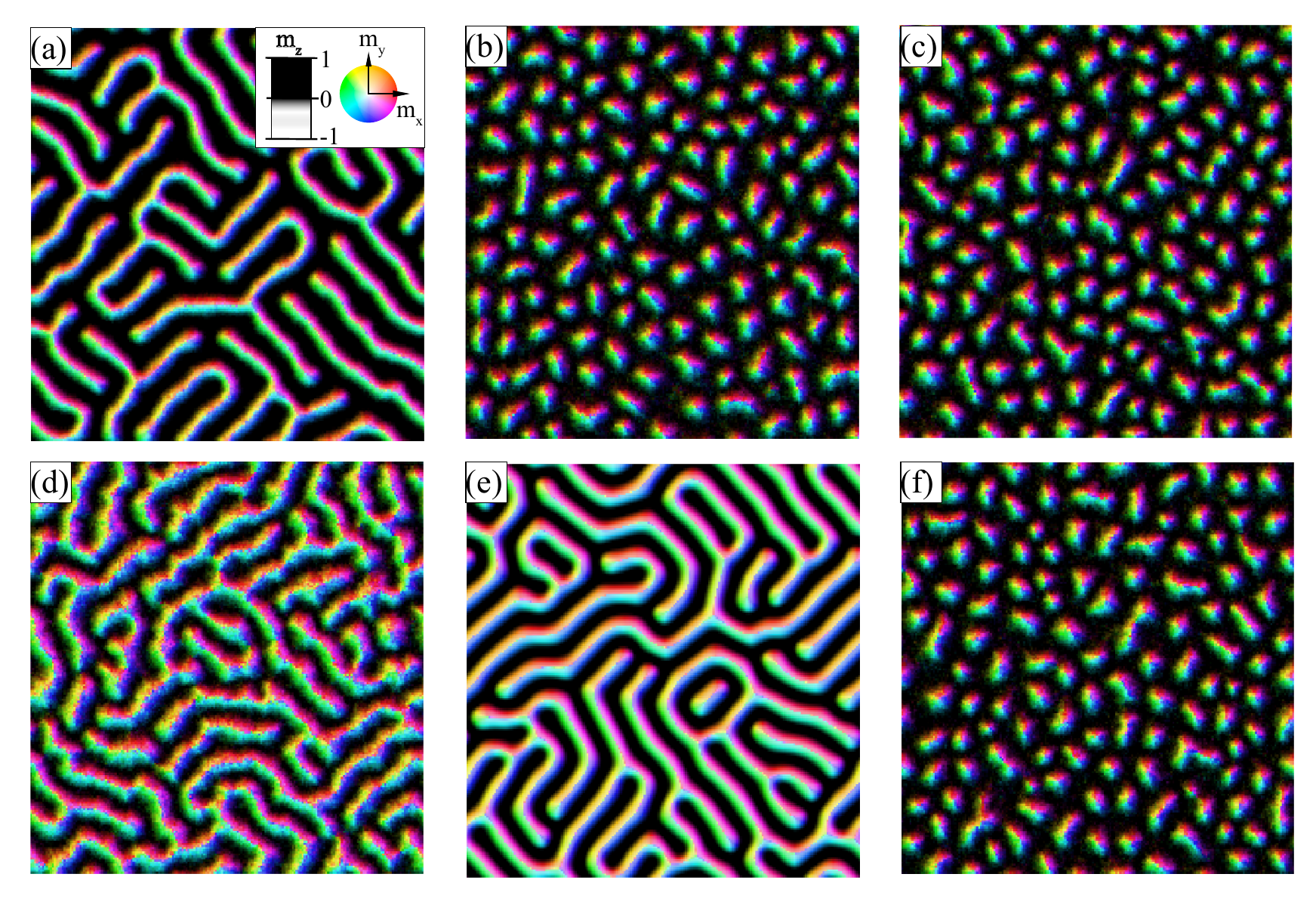}
	\caption{(a, b) Structures after $30\,$ns evolution at $\mu_0H=0.3\,$T, and 
		$T=20\,$K (a) and $33\,$K (b), starting from the initial configuration of 
		Fig. \ref{fig10}(c1). (c) Structures after $10$ns evolution at $\mu_0H=0.3\,$T 
		and $T=33\,$K, starting from the initial configuration of Fig.\ref{fig10}(c3).
		(d,e) Structures after $33\,$ns zero field warming (d) and cooing (e), starting 
		from the SkX in Fig. \ref{fig10}(c3). (f) A SkX of 133 skyrmions at $\mu_0H=
		0.3\,$T and $33\,$K comes from the thermally generated nucleation centers.}
	\label{fig12}
\end{figure*}

To substantiate this claim, we use bulk DMI and repeat those simulations 
for Figs. 1, 4, and 6. Figure \ref{fig10} shows the metastable structures 
with 10 (a1-c1), 50 (a2-c2) and 150 (a3-c3) skyrmions under magnetic 
field of 0T (a1-a3), 0.1T (b1-b3) and 0.4T (c1-c3). Similar to Fig. 
\ref{fig1}, the film is in a helical state consisting of ramified stripe 
skyrmions with a small skyrmion number, and a SkX in a triangular lattice 
with a large skyrmion number. $Q_m$ are also the same as those in Fig. 
\ref{fig4} at various magnetic field, although the energy curves are 
slightly different as shown in Fig.~\ref{fig11}. 
To further prove that states with $Q_m$ skyrmions are thermal equilibrium 
states, Figs. \ref{fig12}(a) and \ref{fig12}(b) show the spin structure 
with 11 and 134 skyrmions after $30\,$ns evolution starting from Fig. 
\ref{fig10} (c1) at $20\,$K and $33\,$K under magnetic field of $0.3\,$T. 
Figure \ref{fig12} (c) shows the structure with 132 skyrmions after 10ns 
evolution starting from Fig. \ref{fig12} (c3) at $33\,$K under a magnetic 
field of $0.3\,$T. Figures \ref{fig12}(d) and \ref{fig12}(e) show the spin 
structure with 12 and 2 skyrmions after 33ns zero-field warming and zero-field 
cooling, respectively, starting from \ref{fig10} (c3). The temperature 
change rates are the same as those in the main text. Figure \ref{fig12} (f) 
shows the spin structure with 133 skyrmions after $30\,$ns evolution starting 
from ferromagnetic state at $33\,$K. Clearly, the formation and disappearance 
of SkXs are the same as Fig. \ref{fig6}, the only difference is the 
skyrmion walls change from a Neel type to a Bloch type.

\par 

\par 

\par 

\section{Conclusion and perspectives}
In conclusion, both helical states and SkXs are the collections of skyrmions. 
Stripes, not disks, are the natural shapes of skyrmions when their formation 
energy is negative. The distinct morphologies of helical and SkX states come 
from the skyrmion density. Skyrmions become circular objects and in a triangles 
lattice due to the compression and strong repulsion among highly packed skyrmions. 
Unexpectedly, there are enormous number of metastable states with an arbitrary 
skyrmion density. These metastable states result in the thermodynamic path 
dependence of an actual state that a film is in. The energy of these states 
depends on the skyrmion number and the magnetic field. The role of a magnetic 
field in SkX formation is to create the lowest energy state being a skyrmion 
condensate with a high skyrmion density for $\kappa$ above a critical value. 
Below the critical value, the lowest energy state is still a collection of 
condensed stripes skyrmions even in the optimal field. The possible reason 
for metastability of SkXs in most chiral magnetic film at low temperature is  
due to increase of exchange stiffness constant and magnetic anisotropy such 
that $\kappa$ is below the critical value at low temperature. It is showed that 
our findings are robust to parameter variations as long as $\kappa$ is given. 
Whether the chiral magnets have the interfacial DMI or bulk DMI does not 
change the physics. Of course, the stripes of circular skyrmions will change 
from the Neel type to the Bloch type. We have also numerically shown that the 
SkX entropy is lower than a collection of stripes. In order to realize truly 
stable SkXs at extremely low temperature, one may use those chiral magnets 
whose $\kappa$ is larger than the critical value in the required energy range. 
Our findings have profound implications in skyrmion-based applications. \\ \par

Looking forward, several issues require further studies. According to the current 
findings, the criteria for SkX formation in a chiral magnetic film are $\kappa>2$ 
and a proper perpendicular magnetic field. Although all known materials support 
stable SkX only in a small pocket in the field-temperature plane near the Curie 
temperature to date while SkXs are metastable in low temperatures, it should be 
interesting to search materials that support stable SkXs in low temperatures 
and metastable near the Curie temperature. In fact, there is no reason why one 
cannot have a film that supports stable SkX in all temperatures below the Curie 
temperature under a proper field. The physics around the maximal skyrmion density 
has not been revealed yet in this study. The critical $\kappa$ is a purely 
numerical fact based on three sets of very different materials parameters. 
No clear understanding of its value and related physics has been obtained here. 
In fact, even why a chiral magnetic film prefers a state with a finite skyrmion 
density in a magnetic field prefer is not clear. It again is a purely numerical 
observation at the moment. Thus, we are still in the infant stage of a true 
understanding of SkX formation. 

\section*{Conflicts of interest}
There are no conflicts of interest to declare.\\
\par

\section*{Acknowledgements}
\par
This work is supported by the National Key Research and Development Program 
of China (grant No. 2020YFA0309600), the NSFC (grant No. 11974296), and 
Hong Kong RGC Grants (No. 16301518, 16301619 and 16302321). 
\\ \par



\balance



\end{document}